\documentclass{emulateapj}
\usepackage{graphicx}

\begin{document}

\shorttitle{CLASH Strong-Lensing Analysis of A383}
\shortauthors{Zitrin et al.}

\slugcomment{Submitted to the Astrophysical Journal}

\title{The Cluster Lensing and Supernova Survey with Hubble (CLASH): Strong Lensing Analysis of Abell 383 from 16-Band HST WFC3/ACS Imaging}

\author{A. Zitrin\altaffilmark{1}}
\author{T. Broadhurst\altaffilmark{2,3}}
\author{D. Coe\altaffilmark{4}}
\author{K. Umetsu\altaffilmark{5}}
\author{M. Postman\altaffilmark{4}}
\author{N. Ben\'itez\altaffilmark{6}}
\author{M. Meneghetti\altaffilmark{7}}
\author{E. Medezinski\altaffilmark{8}}
\author{S. Jouvel\altaffilmark{9}}
\author{L. Bradley\altaffilmark{4}}
\author{A. Koekemoer\altaffilmark{4}}
\author{W. Zheng\altaffilmark{8}}
\author{H. Ford\altaffilmark{8}}
\author{J. Merten\altaffilmark{10}}
\author{D. Kelson\altaffilmark{11}}
\author{O. Lahav\altaffilmark{9}}
\author{D. Lemze\altaffilmark{8}}
\author{A. Molino\altaffilmark{6}}
\author{M. Nonino\altaffilmark{12}}
\author{M. Donahue\altaffilmark{13}}
\author{P. Rosati\altaffilmark{14}}
\author{A. Van der Wel\altaffilmark{15}}
\author{M. Bartelmann\altaffilmark{10}}
\author{R. Bouwens\altaffilmark{16}}
\author{O. Graur\altaffilmark{1}}
\author{G. Graves\altaffilmark{17}}
\author{O. Host\altaffilmark{9}}
\author{L. Infante\altaffilmark{18}}
\author{S. Jha\altaffilmark{19}}
\author{Y. Jimenez-Teja\altaffilmark{6}}
\author{R. Lazkoz\altaffilmark{2}}
\author{D. Maoz\altaffilmark{1}}
\author{C. McCully\altaffilmark{19}}
\author{P. Melchior\altaffilmark{20}}
\author{L.A. Moustakas\altaffilmark{21}}
\author{S. Ogaz\altaffilmark{4}}
\author{B. Patel\altaffilmark{19}}
\author{E. Regoes\altaffilmark{22}}
\author{A. Riess\altaffilmark{4,8}}
\author{S. Rodney\altaffilmark{8}}
\author{S. Seitz\altaffilmark{23}}

%\affil{
\altaffiltext{1}{The School of Physics and Astronomy, Tel Aviv University; adiz@wise.tau.ac.il}
\altaffiltext{2}{Department of Theoretical Physics, University of Basque Country}
\altaffiltext{3}{IKERBASQUE, Basque Foundation for Science}
\altaffiltext{4}{Space Telescope Science Institute}
\altaffiltext{5}{Institute of Astronomy and Astrophysics, Academia Sinica}
\altaffiltext{6}{Instituto de Astrof\'isica de Andaluc\'ia (CSIC)}
\altaffiltext{7}{INAF, Osservatorio Astronomico di Bologna; INFN, Sezione di Bologna}
\altaffiltext{8}{Department of Physics and Astronomy, The Johns Hopkins University}
\altaffiltext{9}{University College London}
\altaffiltext{10}{Universitat Heidelberg}
\altaffiltext{11}{Carnegie Institution}
\altaffiltext{12}{INAF-Osservatorio Astronomico di Trieste}
\altaffiltext{13}{Michigan State University}
\altaffiltext{14}{European Southern Observatory}
\altaffiltext{15}{MPIA, Heidelberg}
\altaffiltext{16}{University of Leiden}
\altaffiltext{17}{UC Berkeley}
\altaffiltext{18}{Universidad Catolica de Chile}
\altaffiltext{19}{Rutgers University}
\altaffiltext{20}{The Ohio State University}
\altaffiltext{21}{Jet Propulsion Laboratory, California Institute of Technology}
\altaffiltext{22}{European Laboratory for Particle Physics (CERN)}
\altaffiltext{23}{Universitas Sternwarte Muenchen}
%}

\begin{abstract}

  We examine the inner mass distribution of the relaxed galaxy cluster Abell
  383 ($z=0.189$), in deep 16-band HST/ACS+WFC3 imaging taken as part of the CLASH multi-cycle treasury program. Our program is designed to study the dark matter distribution in 25 massive clusters, and balances depth with a wide wavelength coverage, 2000--16000\AA, to
  better identify lensed systems and generate precise photometric redshifts. This photometric information together with
  the predictive strength of our strong-lensing analysis method identifies 13 new
  multiply-lensed images and candidates, so that a total of 27 multiple-images of 9 systems are used to tightly constrain the inner
  mass profile gradient, $d\log \Sigma/d\log r\simeq -0.6\pm 0.1$
  ($r<160$kpc). We find consistency with the standard
  distance-redshift relation for the full range spanned by the lensed
  images, $1.01<z<6.03$, with the higher redshift sources deflected
  through larger angles as expected. The inner mass profile derived
  here is consistent with the results of our independent weak-lensing
  analysis of wide-field Subaru images, with good agreement in the
  region of overlap ($\sim0.7-1$ arcmin). Combining weak and strong lensing, the
  overall mass profile is well fitted by an NFW profile with $M_{vir}=(5.37^{+0.70}_{-0.63}\pm 0.26) \times 10^{14}M_{\odot}/h$ and a
  relatively high concentration, $c_{vir} = 8.77^{+0.44}_{-0.42}\pm 0.23$,
  which lies above the standard $c$--$M$ relation similar to other well-studied clusters. The critical radius of Abell 383 is modest by the standards of other lensing clusters, $r_{E}\simeq16\pm2\arcsec$ (for $z_s=2.55$), so the relatively large
  number of lensed images uncovered here with precise photometric
  redshifts validates our imaging strategy for the CLASH survey. In
  total we aim to provide similarly high-quality lensing data for 25 clusters, 20 of which are X-ray selected relaxed clusters, enabling a precise determination of the representative mass profile free from lensing bias.
\end{abstract}

%\begin{keywords}
%\end{keywords}

\section{Introduction}\label{intro}

Clusters of galaxies play a direct and fundamental role in testing
cosmological models and in constraining the properties of dark matter
(DM), providing unique and independent tests of any viable cosmology
and structure formation scenario \citep[e.g.,][]{Lahav1991, Evrard2002, Broadhurst2005b, Lemze2009, Jullo2010}. Their extreme virial masses mean that unlike individual galaxies, gas cooling is not capable of
compressing the dark matter halo, so that cluster mass profiles
reflect directly the thermal evolution of the DM and the growth of the
cosmological density field \citep{Peebles1985, Duffy2010}. The capability of clusters to
critically examine the standard cosmological model is now welcomed
more than ever given the unattractive hybrid nature of the standard
$\Lambda$CDM model derived by other means.

Simulated CDM dominated halos consistently predict mass profiles that
steepen with radius, providing a distinctive, fundamental prediction
for this form of DM (\citet{Navarro1996}; NFW).
Furthermore, the degree of mass concentration should decline with
increasing cluster mass because clusters that are more massive,
collapse later, when the cosmological background density is lower
\citep[e.g.,][]{Bullock2001, Zhao2003, Neto2007}. Cluster lensing provides
a model independent means of testing these fundamental predictions. Given an unbiased sample of relaxed clusters with high spatial
resolution, one can rigorously test these basic predictions of the
standard $\Lambda$CDM model and contending scenarios. To date, only
limited progress has been made toward these aims given the
considerable observational challenges of obtaining data of sufficient
quality for accurate weak and strong lensing work.

Full mass profiles spanning the weak and strong lensing regimes have
been constructed for only a handful of clusters, involving deep HST data to reliably
identify large samples of multiple images, and high quality wide-field
imaging for careful weak-lensing (WL) work \citep[e.g.,][]{Gavazzi2003, Broadhurst2005b, Broadhurst2008, UmetsuBroadhurst2008, Merten2009, Merten2011, Newman2009, Coe2010, Umetsu2010, Umetsu2011a, Zitrin2010}. It has become clear that the
inner mass profile can be accurately obtained using several sets of
multiple images spanning a wide range of redshifts \citep{Zitrin2009b, Zitrin2010, Zitrin2011b}. In
the case of WL the data are readily invertible to obtain a
model-independent mass profile \citep{KaiserSquires1993}, but much
published work has suffered from a significant dilution of the lensing
signal by foreground objects and cluster members, leading to shallow profiles with
underestimated Einstein radii. The ability of multi-color
photometry to isolate foreground and background with reference to the
radial WL signal has been demonstrated by \citet{Medezinski2010}, so that the WL signal is found to be higher than earlier work, particularly so towards the center of the cluster.

The initial results from combining deep strong-lensing (SL) work with minimally-diluted
WL analyses has led to intriguing results, in the sense that
although the mass profiles are well fitted by NFW-like profiles,
showing the continuously steepening logarithmic gradient consistent
with the expected form for CDM dominated halos, the concentration of
matter in these halos seems to lie above the
mass-concentration relation predicted by the standard $\Lambda$CDM model
\citep{Gavazzi2003, Broadhurst2005b, Zitrin2010, Umetsu2011a}. Lensing bias is an
issue here for clusters which are primarily selected by their lensing
properties, where the major axis of a cluster may be aligned
preferentially close to the line of sight, boosting the projected mass
density observed \citep[e.g.,][]{Hennawi2007, CorlessKing2009, OguriBlandford2009, Sereno2010, Morandi2011}. This will usually result also in higher measured concentrations and larger Einstein radii \citep[e.g.,][]{SadehRephaeli2008, Meneghetti2010a}, though even with these effects taken into account there seems to be some discrepancy from $\Lambda$CDM predictions \citep{Oguri2009, Meneghetti2011, Zitrin2011a}. While existing data may not support a strong conclusion that the observations are in significant tension with the standard $\Lambda$CDM model, it is clear that a larger X-ray selected sample, with minimal
lensing bias and excellent SL and WL data, is required to evaluate the significance of these trends.

Several examples of high-redshift virialized clusters with diffuse
X-ray emission are known, where the highest-redshift cluster selected by X-ray
means is now established at $z=2.07$ (CL J1449+0856; \citealt{Gobat2011}). The most massive
of these clusters is XMMU J2235.3-2557 at $z=1.39$ \citep{Rosati2009} with an estimated
total mass of $M_{tot}(< 1Mpc) = (5.9 \pm 1.3) \times 10^{14} M_{\odot}$. The existence of these clusters, as well as the existence of evolved galaxies at high
redshift, are claimed to be unlikely given the predicted abundance of extreme perturbations of
cluster sized masses in the standard $\Lambda$CDM scenario \citep[e.g.,][]{Daddi2007, Daddi2009, Collins2009, Jee2009, Richard2011}, pointing towards a more extended early history of growth, or a non-Gaussian distribution of massive perturbations.

To shed new light on these mysteries we have embarked on a major project
involving galaxy clusters, the \emph{Cluster Lensing And Supernova survey with
Hubble} (CLASH). For more details see \citet{Postman2011CLASHoverview}. The CLASH program has been awarded
524 orbits of HST time to conduct a multi-cycle program that will
couple the gravitational-lensing power of 25 massive intermediate
redshift galaxy clusters with HST's newly enhanced panchromatic
imaging capabilities (WFC3 and the restored ACS), in order to test structure
formation models with unprecedented precision. The CLASH observations, combined with our wide-field optical and X-ray
imaging, represent a substantial advance in the quality and
quantity of SL data, enabling us to measure the dark
matter mass profile shapes and mass concentrations from hundreds
of multiply-imaged sources, providing precise ($\sim10\%$) observational
challenges to scenarios for the DM mass distribution (for full details about the CLASH program see \citealt{Postman2011CLASHoverview}).

The 16 HST bands chosen for this project ranging from the UV through the
optical and to the IR, and additional spectra available from large ground-based
telescopes for some of the brighter arcs, enable us to obtain accurate redshifts for the multiply-lensed sources presented in this work. We use these remarkable imaging data along with our well-tested approach
to SL modeling \citep[e.g.,][]{Broadhurst2005a, Zitrin2009a, Zitrin2009b, Zitrin2010, Zitrin2011a, Zitrin2011b}, in order to find a significant number of multiple images
across the central field of Abell 383 (A383 hereafter) so that its mass distribution
and profile can be constrained with high precision. Various other
mass models for this cluster were previously presented \citep[e.g.,][]{Smith2001, Smith2005, Sand2004, Sand2008, Newman2011} usually based on
WFPC2/HST single-band observations, uncovering 3-4 multiple image-systems
and various candidates, as will be further discussed in \S
\ref{Mimages}.

The approach to SL modeling implemented here involves only six free parameters so
that in practice the number of multiple images uncovered readily
exceeds the number of free parameters as minimally required in order
to obtain a reliable fit, allowing for identification of other
multiply-lensed systems across the cluster field.  Our approach to
lens-modeling is based on the reasonable assumption that mass
approximately traces light. We have independently tested
this assumption in Abell 1703 \citep{Zitrin2010}, by applying the
non-parametric technique of \citet{Liesenborgs2006, Liesenborgs2007, Liesenborgs2009} for
comparison, yielding similar results. Such parameter-free
methods usually do not have the precision to actually find new multiple-images,
but the resulting 1D radial profiles are sufficiently accurate for meaningful
comparisons. Independently, it has been found
that SL methods based on parametric modeling are accurate at the
level of a few percent in determining the projected inner mass
\citep{Meneghetti2010b}.

The paper is organized as follows: In \S 2 we describe the
observations, and in \S 3 we detail the SL analysis. In \S 4 we report
and discuss the results where in \S 5 we compare these to numerical simulations. The results are then summarized in \S6. Throughout
this paper we adopt a concordance $\Lambda$CDM cosmology with
($\Omega_{\rm m0}=0.3$, $\Omega_{\Lambda 0}=0.7$, $h=0.7$). With these
parameters one arcsecond corresponds to a physical scale of 3.17 kpc
for this cluster (at $z=0.189$; \citealt{Sand2004}). The reference center of our analysis
is fixed on the brightest cluster galaxy (BCG): RA = 02:48:03.41 Dec = -03:31:44.91
(J2000.0).

\section{Observations and Redshifts}\label{obs}

As part of the CLASH program (see \S1),
Abell 383 was observed with {\em HST} between 2010 November to 2011 March.
This is our first of 25 clusters to be observed to a depth of 20 {\em HST} orbits in 16 filters
with the Wide Field Camera 3 (WFC3) UVIS and IR cameras,
and the Advanced Camera for Surveys (ACS) WFC.
Observation details and filters are provided in Table \ref{exp}.

The images are processed for debias, flats, superflats, and darks, using standard techniques.
The ACS images are further corrected for bias striping \citep{Grogin2010}
% http://www.stsci.edu/hst/acs/software/destripe/destriping_hstcal2010.pdf
and CTE/CTI degradation effects \citep{AndersonBedin2010}.
% http://adsabs.harvard.edu/abs/2010PASP..122.1035A
WFC3/IR pixels are flagged and downweighted for persistence effects.
All images are then co-aligned and combined using drizzle algorithms
to a scale of $0.065\arcsec /$ pixel. An additional set of images with the original ACS $0.05\arcsec /$ pixel scale is produced, onto which we apply our modeling initially to maintain the higher resolution, where the full UVIS/ACS/WFC3-IR data set is then importantly used for multiple-images verification and measurement of their photometric redshifts. Further details of our pipeline will be presented in an upcoming paper.

Based on the 16-filter photometry,
we obtain photometric redshifts using BPZ \citep{Benitez2000, Benitez2004, Coe2006}
and LePhare (LPZ hereafter; \citealt{ArnoutsLPZ1999,Ilbert2006BPZ}).
These two methods yielded some of the best results
of all photo-z methods tested by the PHoto-z Accuracy Testing group \citep{Hildebrandt2010PHOTZ}.
BPZ and LPZ are similar in that spectral energy distribution (SED) templates
are redshifted and fit to observed photometry.
BPZ currently uses 6 templates from PEGASE \citep{Fioc1997PEGASE},
calibrated using the FIREWORKS photometry and spectroscopic redshifts from \citet{Wuyts2008PHOTZ}.
LPZ uses templates from \citet{BC03}
calibrated using COSMOS \citep{Koekemoer2007} photometry and spectroscopic redshifts
as described in \citet{Ilbert2009}. The templates are empirically generated, describing well the full range of galaxy colors found in these multiband catalogs (less than $\sim1\%$ outliers for high quality spectroscopic samples), and therefore implicitly encompass all the range of metallicities, extinctions, and star formation histories of real galaxies. Further details on these methods can be found in the aforementioned references.
While similar, the two methods serve as important cross-checks of one another.

The distances to the galaxies are, of course, key ingredients to the lens model.
The photo-z analyses used here also clearly aid us in assessing the robustness of the multiple-image identifications. The photometry of some lensed images may be significantly contaminated by brighter nearby cluster galaxies.  SExtractor attempts to correct for this by measuring and subtracting the local background around each object.  This works well in some cases but not all.  To better reveal these lensed galaxies, we have carefully modeled and subtracted the light of several cluster galaxies including the BCG.  While this improved the detection of some lensed galaxies, it did not consistently improve their photometry and thus photometric redshifts.  The cluster galaxy wings must be modeled and subtracted very robustly and consistently to achieve quality photometry in all 16 bands for faint, nearby galaxy images.

Explicitly, for its subtraction, the BCG has been modeled using the CHEF basis \citep{Jimenez&Benitez2011}. This
basis comprises both Chebyshev rational and trigonometric functions,
ensuring that the extended disk of this object is properly modeled. The
flexibility of the CHEFs scale parameter allows us to accurately represent
the BCG while keeping significant substructure and
arcs unchanged. An example of the BCG subtraction is seen in Figure \ref{curves383}.

Our 16 filters were selected based on tests with simulated photometry
to yield precise ($\Delta z \sim 0.02(1+z)$) photo-z's \citep{Postman2011CLASHoverview}.
Previous work has also demonstrated how photo-z precision improves
by increasing the number of (preferably overlapping) filters for a fixed total observing time \citep{Benitez2009PHOTZ}.
The empirical precision of CLASH photo-z's for arcs and other galaxies,
including the relative contributions of various filters, will be detailed in future work.

\begin{table}
\caption{CLASH HST Observation Log for A383}
%\vspace{0.5cm}
\label{exp}
%\begin{footnotesize}
\begin{center}
\begin{tabular}{|c|c|c|c|c|}
  \hline\hline
  Filter & Assigned orbits &Total time (s) & Instrument\\
  \hline
  F225W& 1.5 & 3672 & WFC3/UVIS\\
  F275W& 1.5 & 3672 & WFC3/UVIS\\
  F336W& 1.0 & 2434 & WFC3/UVIS\\
  F390W& 1.0 & 2434 & WFC3/UVIS\\
  F435W& 1.0 & 2125 & ACS/WFC\\
  F475W& 1.0 & 2064 & ACS/WFC\\
  F606W& 1.0 & 2105& ACS/WFC\\
  F625W& 1.0 & 2064 & ACS/WFC\\
  F775W& 1.0 & 2042 & ACS/WFC\\
  F814W& 2.0 & 4243 & ACS/WFC\\
  F850LP& 2.0 & 4214 & ACS/WFC\\
  F105W& 1.1 & 2815 & WFC3/IR\\
  F110W& 1.0 & 2515 & WFC3/IR\\
  F125W& 1.0 & 2515 & WFC3/IR\\
  F140W& 1.0 & 2412 & WFC3/IR\\
  F160W& 2.0 & 5029& WFC3/IR\\
  \hline
\end{tabular}
\tablecomments{Observation were carried out between 2010, November 18th, to 2011, March 3rd. The table summarizes the total exposure time in each filter. Note that these values are specific to A383. Observation times may vary for other CLASH clusters. We also note that the $5\sigma$ limiting magnitude is fainter than 26.8 AB mag for all 16 filters, as will be detailed in an upcoming paper \citep{Postman2011CLASHoverview}.}
\end{center}
\end{table}

\section{Strong Lensing Modeling and Analysis}\label{model}

We apply our well tested approach to lens modeling, which has
previously uncovered large numbers of multiply-lensed galaxies in ACS
images of Abell 1689, Cl0024, 12 high-$z$ MACS clusters, MS 1358, and the ``Pandora cluster'' Abell 2744
\citep[respectively,][]{Broadhurst2005a, Zitrin2009b, Zitrin2011a, Zitrin2011b, Merten2011}.
Briefly, the basic assumption adopted is that mass approximately
traces light, so that the photometry of the red cluster member
galaxies is used as the starting point for our model. Cluster member
galaxies are identified as lying close to the cluster sequence by the
photometry described in \S \ref{obs}. In addition, using our extensive 16-band imaging and corresponding photometric redshifts, these can be then verified as members lying at the cluster's redshift.

\begin{figure}
 \begin{center}
   \includegraphics[width=84mm,height=80mm]{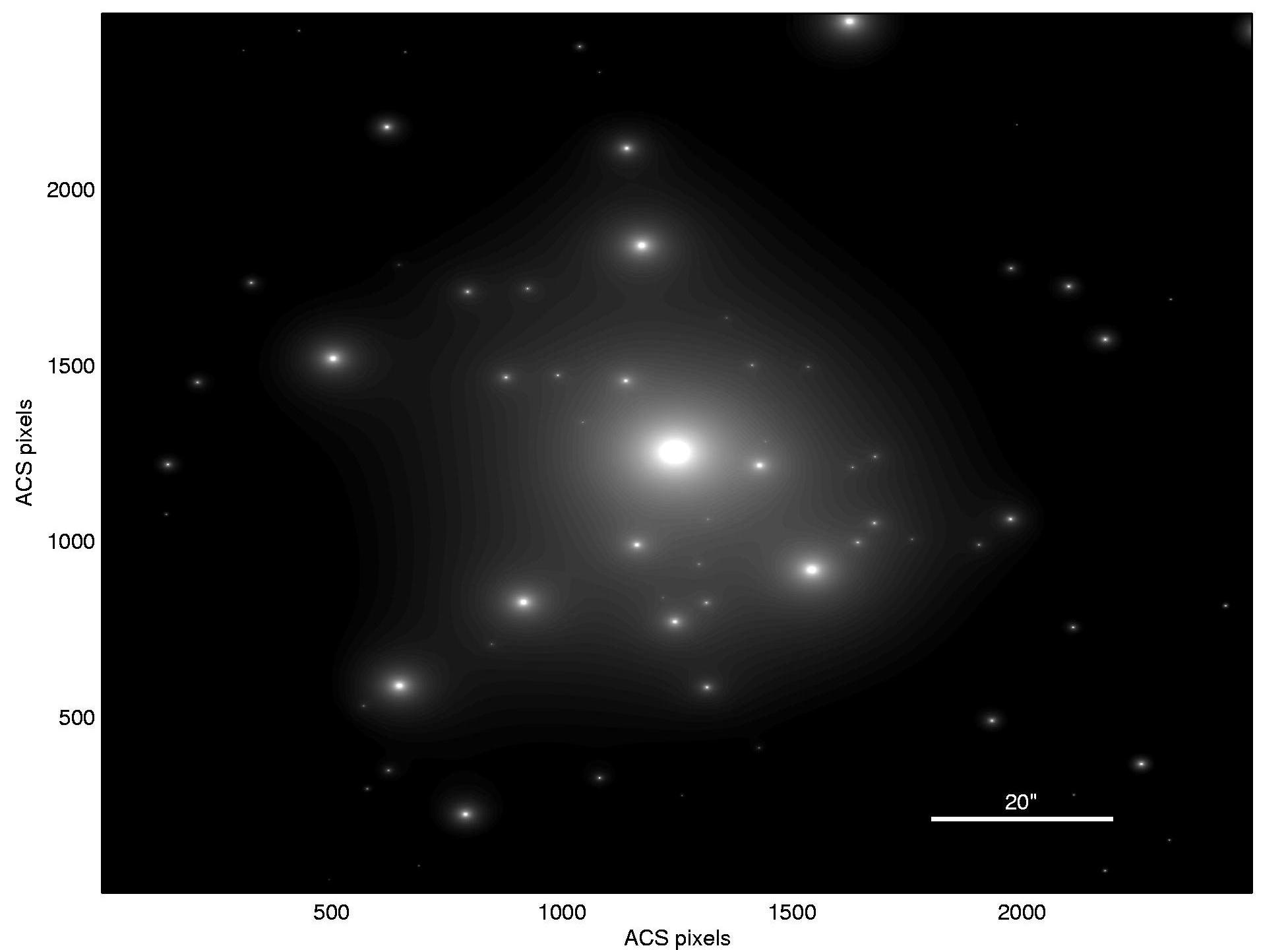}
   \end{center}
\caption{The starting point of the mass model, where we define the
surface mass distribution based on the cluster member galaxies (see \S \ref{model}). Axes are in ACS pixels ($0.05 \arcsec /pixel$), and a $20\arcsec$ bar is overplotted. North is up, East is left.}
\label{lumpyhcomp}
\end{figure}

\begin{figure}
 \begin{center}
   \includegraphics[width=85mm]{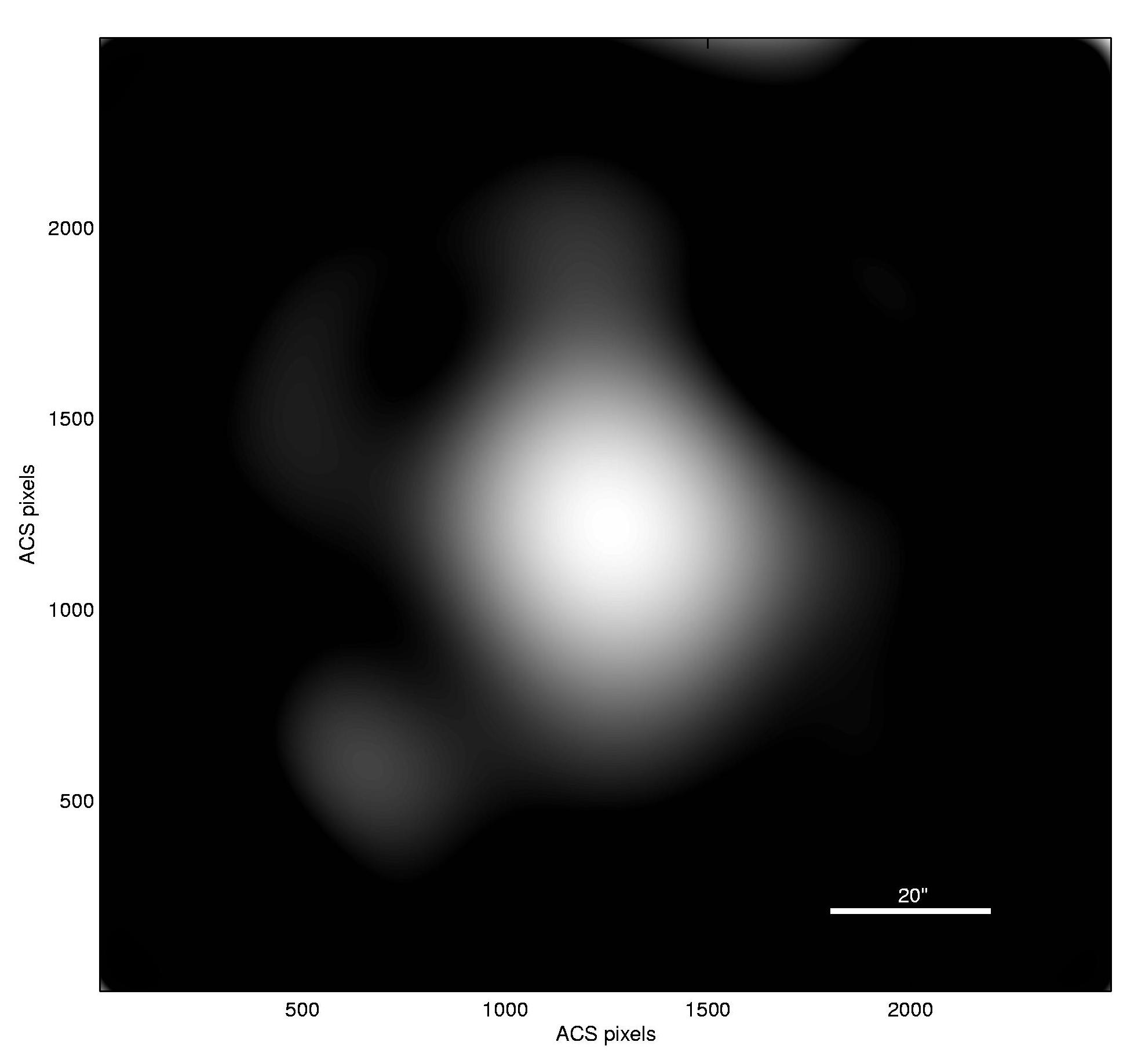}
 \end{center}
\caption{The resulting smooth mass component of the mass model (see \S \ref{model}). Axes are in ACS pixels ($0.05 \arcsec /pixel$), and a $20\arcsec$ bar is overplotted. North is up, East is left.}
\label{smoothcomp}
\end{figure}

 We approximate the large scale distribution of cluster mass by assigning a
 power-law mass profile to each galaxy (see Figure \ref{lumpyhcomp}), the
 sum of which is then smoothed (see Figure \ref{smoothcomp}). The
 degree of smoothing ($S$) and the index of the power-law ($q$) are
 the most important free parameters determining the mass profile. A worthwhile improvement in fitting the location of the lensed images
 is generally found by expanding to first order the gravitational
 potential of this smooth component, equivalent to a coherent shear
 describing the overall matter ellipticity. The direction of the
 shear ($\phi_{\gamma}$) and its amplitude ($|\gamma|$) are free parameters, allowing for some flexibility in
 the relation between the distribution of DM and the distribution of
 galaxies, which cannot be expected to trace each other in detail. The total deflection field $\vec\alpha_T(\vec\theta)$, consists of the
 galaxy component, $\vec{\alpha}_{gal}(\vec\theta)$, scaled by a
 factor $K_{gal}$, the cluster DM component
 $\vec\alpha_{DM}(\vec\theta)$, scaled by (1-$K_{gal}$), and the
 external shear component $\vec\alpha_{ex}(\vec\theta)$, all scaled by the overall normalization factor $K_{q}$:

\begin{equation}
\label{defTotAdd}
\vec\alpha_T(\vec\theta)= K_{q} (K_{gal} \vec{\alpha}_{gal}(\vec\theta)
+(1-K_{gal}) \vec\alpha_{DM}(\vec\theta)
+\vec\alpha_{ex}(\vec\theta)),
\end{equation}
where the deflection field at position $\vec\theta_m$
due to the external shear,
$\vec{\alpha}_{ex}(\vec\theta_m)=(\alpha_{ex,x},\alpha_{ex,y})$,
is given by:
\begin{equation}
\label{shearsx}
\alpha_{ex,x}(\vec\theta_m)
= |\gamma| \cos(2\phi_{\gamma})\Delta x_m
+ |\gamma| \sin(2\phi_{\gamma})\Delta y_m,
\end{equation}
\begin{equation}
\label{shearsy}
\alpha_{ex,y}(\vec\theta_m)
= |\gamma| \sin(2\phi_{\gamma})\Delta x_m -
  |\gamma| \cos(2\phi_{\gamma})\Delta y_m,
\end{equation}
where $(\Delta x_m,\Delta y_m)$ is the displacement vector of the
position $\vec\theta_m$ with respect to a fiducial reference position,
which we take as the lower-left pixel position $(1,1)$, and
$\phi_{\gamma}$ is the position angle of the spin-2 external
gravitational shear, measured counter-clockwise from the $x$-axis. The normalization of the model ($K_{q}$) and the relative scaling of the smooth DM
component versus the galaxy contribution ($K_{gal}$) bring the total number of
free parameters in the model to 6 (see \citealt{Zitrin2009b} for more details). This approach to SL is sufficient
to accurately predict the locations and internal structure of multiple
images, since in practice the number of multiple images uncovered
readily exceeds the number of free parameters, so that the fit is fully constrained.

\begin{figure*}
 \begin{center}
  \includegraphics[width=170mm]{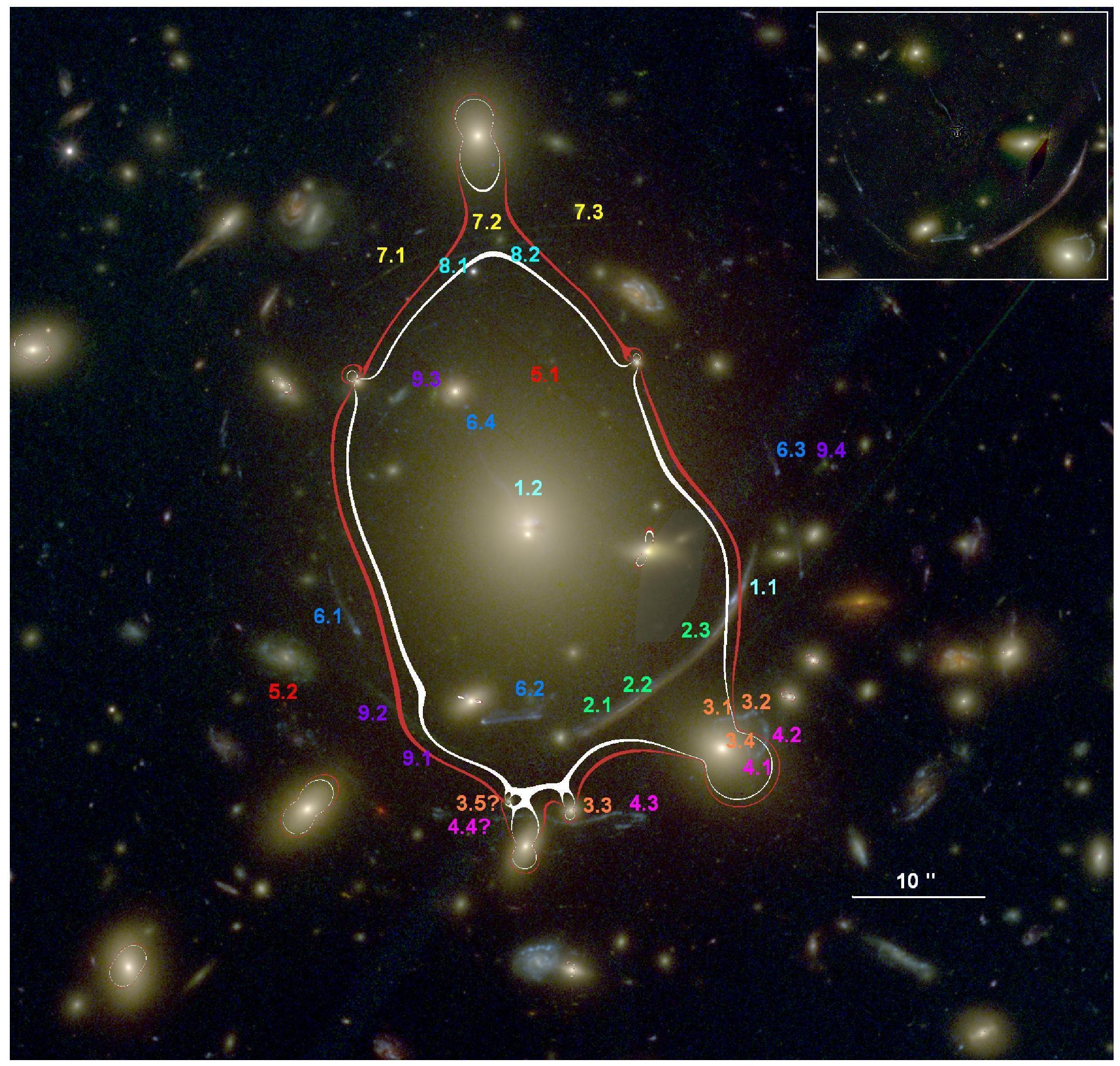}
 \end{center}
\caption{Galaxy cluster A383 ($z=0.189$) imaged
with HST/ACS/WFC3. North is up, East is left. We number the multiply-lensed images used and
uncovered in this work. The numbers indicate the 27 lensed images, 13 of which correspond to 4 newly identified sources, and the different colors are used to
distinguish the 9 different sources. For more details on the each system and the robustness of the new identifications see \S \ref{Mimages}. The overlaid white critical curve corresponds to systems 3 and 4, at $z_{s}=2.55$, enclosing a critical area of an effective Einstein
radius of $\simeq 52$ kpc at the redshift of this cluster ($16.3\arcsec$). Also plotted is a red critical curve, which corresponds to system 5, the drop-out high redshift galaxy at $z_{s}=6.027$. The composition of this color image is Red=F105W+F110W+F125W+F140W+F160W, Green=F606W+F625W+F775W+F814W+F850LP, and Blue=F435W+F475W. This image was generated automatically by using the freely available Trilogy software.\footnote{\href{http://www.stsci.edu/~dcoe/trilogy/}{http://www.stsci.edu/$\sim$dcoe/trilogy/}} The upper-right inset shows the central core with the BCG subtracted, using the method of \citet[][see also \S \ref{obs}]{Jimenez&Benitez2011}.}
\label{curves383}
\end{figure*}

In addition, two of the six free parameters, namely the galaxy power law index $q$, and the smoothing degree $S$, can be primarily set to
reasonable values so that only 4 of the free parameters have to be constrained
initially, which sets a very reliable starting-point using obvious or known
systems. This is because these two parameters control the mass slope, but the overall mass distribution and corresponding critical curves do not strongly depend on them. The mass distribution is therefore primarily well
constrained, uncovering many multiple-images which can then be
iteratively incorporated into the model, by using their redshift
estimation and location in the image-plane.

We use this preliminary model to delens the more obvious lensed
galaxies back to the source plane by subtracting the derived
deflection field. We then relens the source plane in order to predict the
detailed appearance and location of additional counter images, which
may then be identified in the data by morphology, internal structure
and color. The best fit is
assessed by the minimum $\chi^{2}$ uncertainty in the image plane:
\begin{equation} \label{RMS}
\chi^{2}=\sum_{i} ((x_{i}^{'}-x_{i})^2 + (y_{i}^{'}-y_{i})^2) ~/ ~\sigma^{2},
\end{equation}
where $x_{i}^{'}$ and $y_{i}^{'}$ are the locations given by the
model, $x_{i}$ and $y_{i}$ are the real image locations, $\sigma$ is
the error in the location measurement (taken as $0.5\arcsec$), and the sum is over all $N$
images. The model location of each image is the averaged location
given by relensing all other images of the same system. The best-fit
solution is unique in this context, and the model uncertainty is
determined by the location (of predicted images) in the image-plane
itself. Importantly, this image-plane minimization does not suffer
from the bias involved with source-plane minimization,
where solutions are biased by minimal scatter towards shallow mass
profiles with correspondingly higher magnification.

The model is successively refined as additional sets of multiple
images are incorporated to improve the fit, importantly using also
their redshift information for better constraining the mass slope. The mass profile is coupled to the redshift distribution of the different systems, since for each redshift the enclosed mass and correspondingly the deflection angle, depend on the lens and source angular-diameter distances ($D_{l}$, $D_{s}$, respectively). Explicitly,   the deflection angle is defined as $\alpha(\theta)= \frac{4GM(<\theta)}{c^2\theta}\frac{d_{ls}}{d_{s}d_{l}}$, and since the lens distance is constant, the mass slope is constrained through the cosmological relation of the $D_{ls}/D_{s}$ growth with source redshift, where $D_{ls}$ is the distance between the lens and the source. This is seen more clearly in Figure \ref{dlsds}.

\section{Results and Discussion}

\subsection{Multiple-Images, Mass Model and Critical Curves}\label{Mimages}

\begin{table*}
  \caption{Multiple-image systems}
\label{systems}
\begin{center}
\begin{tabular}{|c|c|c|c|c|c|c|c|}
\hline\hline
ARC & RA & DEC & BPZ $z_{phot}$& LPZ $z_{phot}$ & spec-$z$& $z_{model}$ &Comment\\
ID& (J2000.0)&(J2000.0)& (best) [95\% C.L.]& (best) [99\% C.L.] & & & \\
\hline
1.1 & 02:48:02.331& -03:31:49.72&0.97 [0.83--1.05]& 0.93 [0.92--0.93]&1.01& (1.01)&\\
1.2 & 02:48:03.525& -03:31:41.85&0.53 [0.34--0.59]&0.47 [0.47--0.48]&1.01& " & radial image in BCG halo\\%14 filters
\hline
2.1 & 02:48:02.947& -03:31:58.95&0.95 [0.87--1.03]&0.90 [0.90--0.92] &1.01& (1.01) &\\
2.2 & 02:48:02.852& -03:31:58.04&0.96 [0.82--1.04]&0.85 [0.78--0.93]&1.01& "&\\
2.3 & 02:48:02.452& -03:31:52.84&0.84 [0.77--0.91]&0.76 [0.67--0.84]&(1.01)& "&\\
\hline
3.1 & 02:48:02.426& -03:31:59.40 &2.79 [2.64--2.94]&2.90 [2.54--3.15]&2.55& (2.55)&\\
3.2 & 02:48:02.309 & -03:31:59.21 &2.90 [2.75--3.05]&3.01 [2.92--3.08]&2.55& "&\\
3.3 & 02:48:03.026 & -03:32:06.75 &2.56 [2.42--2.70]&3.03 [2.86--3.09]&2.55& "&\\
3.4 & 02:48:02.300 & -03:32:01.74 &2.88 [2.73--3.05]&3.01 [2.87--3.16]&(2.55)& "&\\
\hline
4.1 &02:48:02.244 &-03:32:02.07 &0.20 [0.15--0.25]&0.20 [0.20--0.26]&2.55& (2.55)&\\%2.79 [2.42--3.16]14 filters %16 give 0.20 [0.15--0.25]
4.2 &02:48:02.214 &-03:32:00.25 &2.85 [2.70--3.00]&2.91 [2.82--3.01]&2.55& "&\\
4.3 &02:48:02.847 &-03:32:06.68&3.09 [2.93--3.25]&3.05 [2.91--3.20]&2.55& "&\\
\hline
5.1 &02:48:03.264 &-03:31:34.77 &5.95 [5.68--6.22]& 5.87 [5.64--5.99]&6.027& (6.027)&\\
5.2 &02:48:04.600 &-03:31:58.47 &6.01 [5.74--6.29]& 5.96 [5.72--6.12]&6.027& "&\\
\hline
%%AZ updated here by Dan
6.1 &02:48:04.272 &-03:31:52.77 &2.67 [2.53--2.81]  &2.13 [2.05--2.18] & --& $\simeq2.0$ & \\
6.2 &02:48:03.377 &-03:31:59.27 &2.38 [2.25--2.51] & 1.93 [1.90--2.09]& --& "& \\
6.3 &02:48:02.153 &-03:31:40.88 &1.89 [1.78--2.04] & 2.10 [1.90--2.20]& --&"& \\%14 filters2.22 [1.90--2.54]%1.92 [1.79--2.21] by 16
6.4 &02:48:03.720 &-03:31:35.87 &1.80 [1.69--1.91] &1.54 [1.46--1.57]& --&"& bright galaxy nearby\\
\hline
7.1 &02:48:04.089 &-03:31:25.54 &4.60 [0.64--4.82]&4.50 [4.24--4.76]&-- &$\simeq4.6$& bimodal\\
7.2 &02:48:03.568 &-03:31:22.55 &4.65 [0.38--5.20]&4.77 [0.52--5.58]& --&"&"\\
7.3 &02:48:03.130 &-03:31:22.16 &4.70 [4.35--5.07]&4.56 [0.20--5.08]&--& "&"\\
\hline
8.1 &02:48:03.681&-03:31:24.43 &0.34 [0.24--2.43]&0.33 [0.20--3.19]& --&$\simeq3.1$&bimodal\\
8.2 &02:48:03.386 &-03:31:23.46 &2.94 [2.41--3.25]&2.93 [0.20--3.46]&-- &"&\\
\hline
%%AZ updated here by Dan
9.1 &02:48:03.920&-03:32:00.83&3.91 [3.63--4.10] &3.83 [3.51--4.09]& --&$\simeq4.0$&\\%14 filters4.28 [3.76--4.80];  0.64 [0.48--0.71] by 16
9.2 &02:48:04.046& -03:31:59.21&0.48 [0.26--0.54]&0.47 [0.39--0.53]&-- &"&segment yields $z_{phot}\sim3.9$, see \S \ref{Mimages}\\
9.3 &02:48:03.872&-03:31:35.03&3.96 [3.77--4.15]&3.57 [3.56--3.59]& --&"&\\
9.4 &02:48:01.918& -03:31:40.23&3.80 [3.61--3.99]&3.75 [3.57--3.82]&--& "&\\
\hline\hline
\end{tabular}
\tablecomments{Multiple-image systems and candidates used and uncovered by our model. For more detailed information on each system see the corresponding subsection. The columns are: arc ID; RA
    and DEC in J2000.0; best photo-$z$ using BPZ, along with 95\% confidence level, minimal and maximal
    photo-$z$; best photo-$z$ using LPZ, along with 99\% confidence level, minimal and maximal
    photo-$z$; spectroscopic redshift, spec-$z$ ; $z_{model}$, estimated redshift for the arcs which lack spectroscopy as predicted by the
    mass model; comments. System 1 was uncovered by \citet{Smith2001,Smith2005} who measured its redshift spectroscopically, which is the
    value given below. Systems 2-5 were also found and spectroscopically measured in previous works \citep{Smith2001, Smith2005, Sand2004, Sand2008, Newman2011, Richard2011}. Note also that unusually large errors in the
    photo-$z$ imply a bimodal distribution. In such cases the values which
    agree with the SL model can be different than specified in the best
    photo-$z$ column, as they arise from another peak in the
    distribution. Such cases are specified in the comments.}
\end{center}
\end{table*}

In addition to the previously-known systems (see \citealt{Newman2011} and references therein, \citealt{Richard2011}), our
modeling technique has uncovered 13 new multiply-lensed images and candidates in the
central field of A383, belonging to 4 new systems. We thus substantially increase in this work the number of available constraints on the mass profile of this cluster.

We have made use of the location and redshift information of the multiple-images to fully constrain the mass model. In our minimization procedure, we obtain for most important parameters controlling the mass distribution, values of $q=1.08\pm0.08$ and $S=12\pm2$, but note these are highly coupled to the photometry used to construct the mass model, and to our procedure detailed in \S 3.

\begin{figure}
 \begin{center}
   \includegraphics[width=80mm]{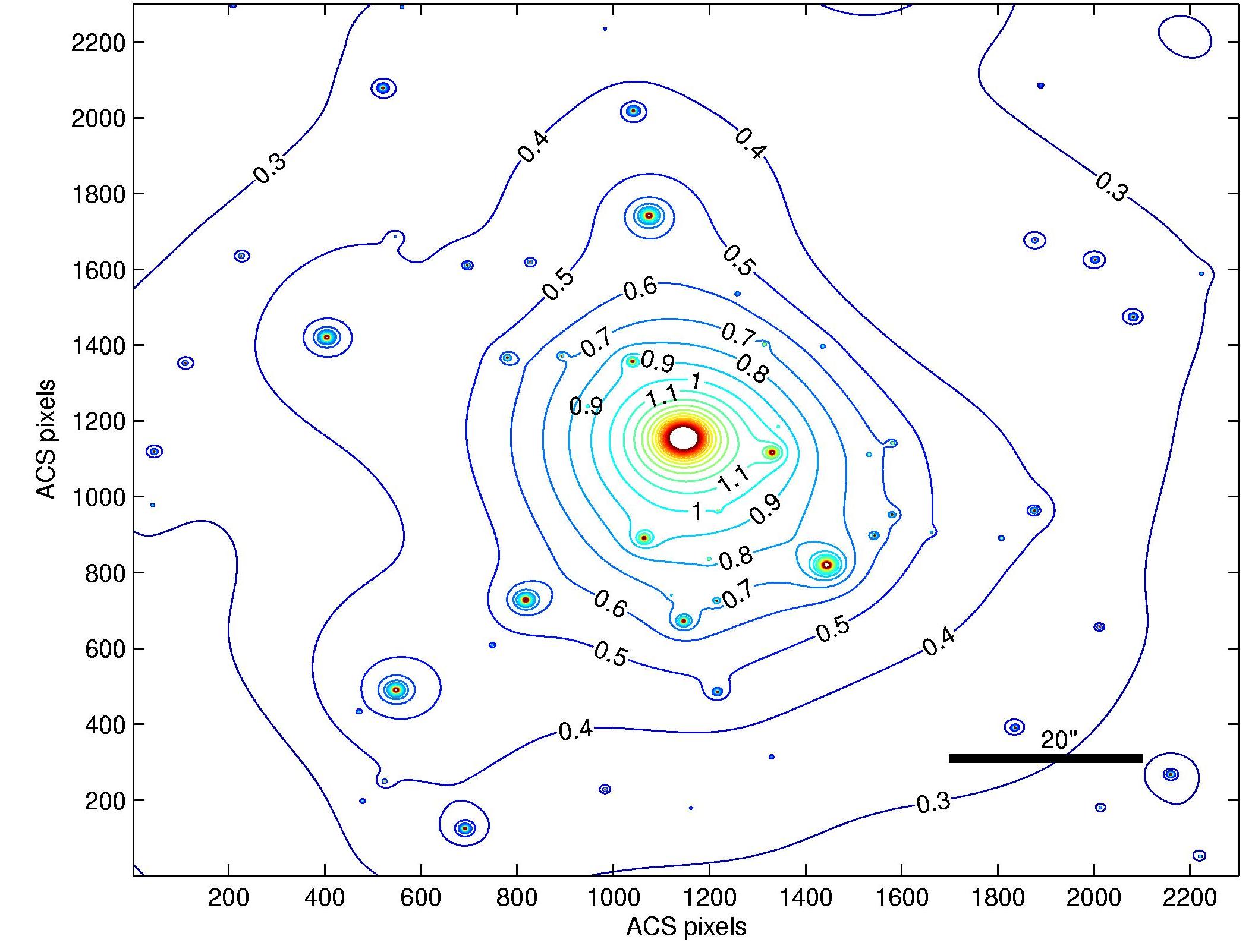}
 \end{center}
\caption{2D surface mass distribution ($\kappa$), in units of the
critical density (for $z_s=2.55$), of A383. Contours are shown in
linear units and in spaces of $\Delta\kappa=0.1$, derived from our mass model constrained using the many
sets of multiply-lensed images seen in Figure
\ref{curves383}. As can be seen, the mass distribution is fairly round. Axes are in ACS pixels ($0.05 \arcsec /pixel$), and a $20\arcsec$ bar is overplotted. North is up, East is left.}
\label{contoursAdi}
\end{figure}

\begin{figure}
 \begin{center}
  \includegraphics[width=85mm]{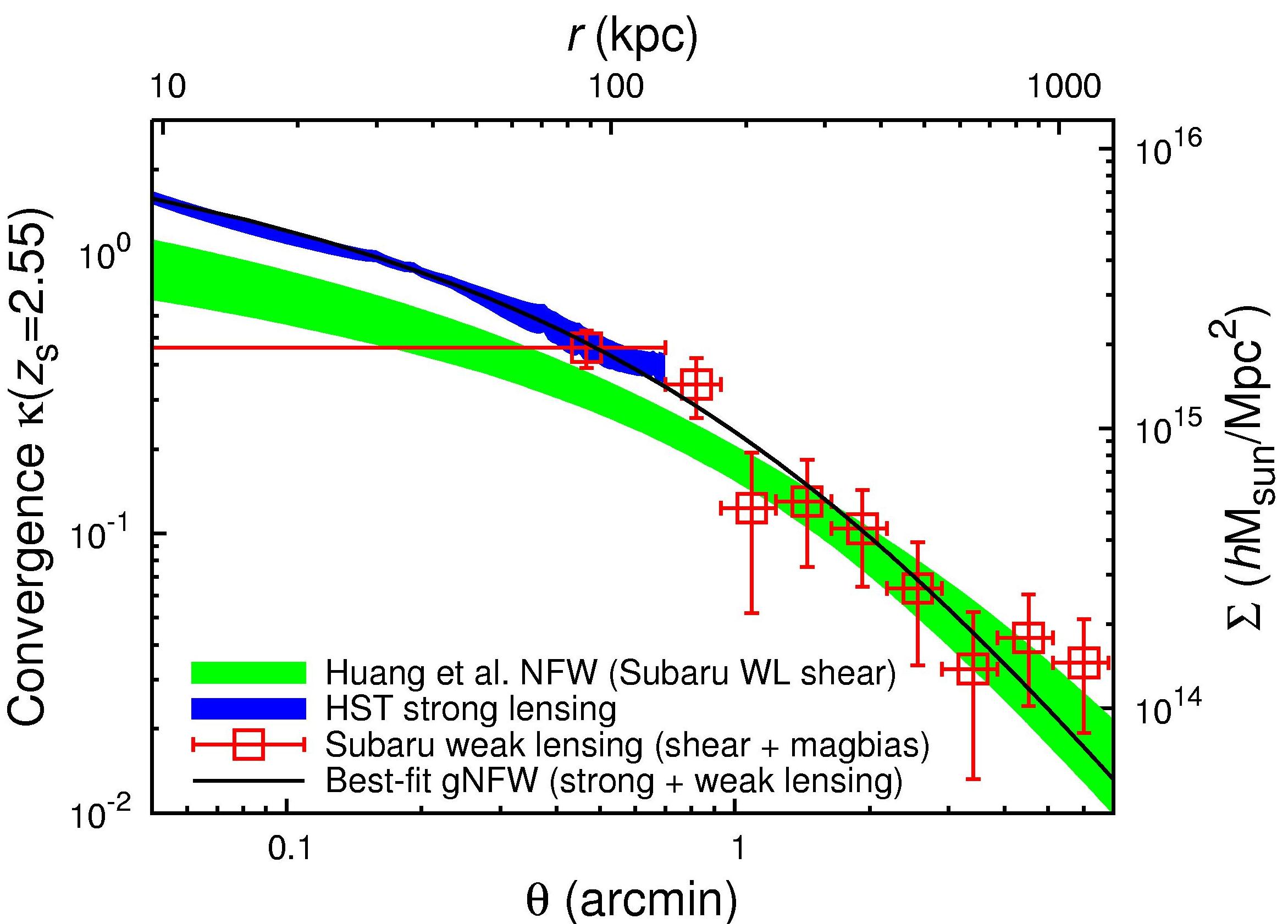}
 \end{center}
\caption{Radial surface mass density ($\kappa$) profile in units of the
critical surface density (for a source redshift of $z_s=2.55$). The inner SL data were
derived using the sets of multiple images shown in Figure
\ref{curves383}. We overplot our preliminary WL data analysis. As can be seen, these are in very good agreement in the region of overlap. A joint SL+WL gNFW fit yields
$M_{vir}=(5.94^{+1.05}_{-0.87}\pm 0.71) \times 10^{14}M_{\odot}/h$
(or $M_{vir}\simeq 8.49 \times 10^{14}M_{\odot}$) and a concentration parameter of
$c_{-2} = 7.95^{+0.89}_{-0.90}\pm 0.55$. The parameter $\alpha=1.078^{+0.069}_{-0.073}\pm 0.059$, so that the overall fit is similar to a simple NFW (see \S \ref{magsec} for explicit comparison). These values are in common with more massive well studied clusters, and lie above the standard $c$--$M$ relation, as seen in Fig. \ref{CM}. Also plotted is the 1D WL analysis of \citet{Huang2011}. A clear consistency is seen through the extensive WL range, though our profile is more consistent with the SL data and is not underestimated in the inner region. A more thorough, 2D WL analysis will be published soon (Umetsu et al., in preparation).}
\label{profileAdi}
\end{figure}

We find that the critical curve for a source
at $z_s=2.55$ (systems 3-4) encloses an area with an effective Einstein
radius of $r_{E}=16.3\pm2\arcsec$, or $\simeq$52 kpc at the redshift
of the cluster. A projected mass of $M= 2.4\pm0.2 \times
10^{13}M_{\odot}$ is enclosed by this critical curve (see Figure
\ref{curves383}). For general comparison, this is in
good agreement with the Einstein radius-mass relation for a source at
$z_s\simeq2-2.5$, found in \citealt{Zitrin2011a} (taking into account also
the different lens distances; see Figure 27 therein). This is naturally expected from the lensing equations, though constitutes an important consistency check. The
corresponding critical curves are plotted on the cluster image in
Figure \ref{curves383} along with the multiply-lensed systems. The
resulting mass distribution and its profile are shown in Figures
\ref{contoursAdi} and \ref{profileAdi}.

It should be stressed that the multiple-images found here are
accurately reproduced by our model and are not simple identifications by
eye. The parametric method of \citet{Zitrin2009b} has been shown in
many cases to have the predictive power to find multiple images in
clusters. Due to the small number of parameters this model is initially well-constrained enabling a reliable identification of other
multiple-images in the field, which can be then used to fine-tune the
mass model. Naturally, the mass model predictions have to be identified in the data and verified further by comparing the SEDs and photometric redshifts of the candidate multiple-images, especially in cases where the images are not prominently bright and big, so that internal details cannot be reliably distinguished. As some of the objects identified here are faint and some may be contaminated by nearby cluster members even after their subtraction, for the less secure cases we supply also the photo-$z$ distributions and spectral energy distributions (SEDs) from our 16 HST-band imaging, so that the reader could assess the plausibility of these identifications. We now detail each multiply-lensed system, as listed
in Table \ref{systems}:

$Systems ~1-2:$ The prominent giant arc, nearly $20\arcsec$ long,
most likely consists of two sources (systems 1 and 2 here) at the same redshift of
$z_{s}=1.01$ \citep{Sand2004, Sand2008, Smith2005}. An
additional radial counter image is seen in the BCG halo (see also
\citealt{Newman2011}). This system was identified by \citet{Smith2001}
in WFPC2 1-band imaging, who also spectroscopically measured the west
side of the arc to be at $z_{s}=1.01$. Following measurement of \citet{Sand2004} with a slit passing through the BCG, the radial arc, and
the eastern part of the main arc, yielded an identical redshift of
$z_{s}=1.01$ for both as well.

\begin{figure}
 \begin{center}
  \includegraphics[width=85mm]{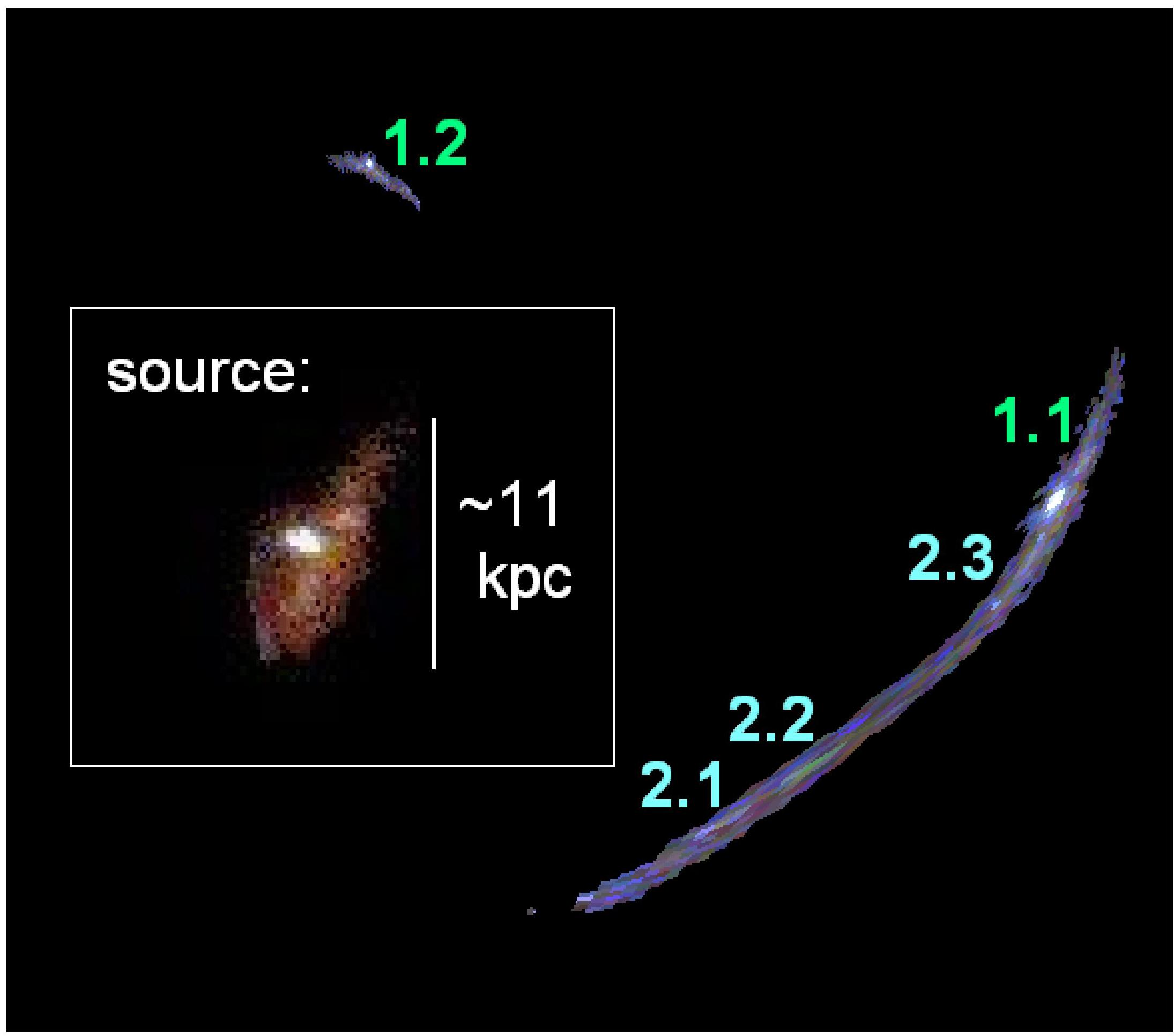}
 \end{center}
\caption{Reproduction of systems 1 and 2 (consisting the giant arc) by our model. We lens the full giant arc to the source plane and back, with a lensing distance corresponding to $z_s=1.01$. Overplotted are the reproduced radial arc, 1.2, and an enlarged image of the reproduced source along with its physical scale. The prominent bright blob lensed in the procedure may be unrelated. The source image had color manipulation and noise cleaning procedures acted on, to better show the internal details.}
\label{rep_sys1}
\end{figure}

Following examples from other well known clusters, it is not common for a giant arc to consist of two different sources. We therefore primarily do not use the location of the multiple-images of these systems in our minimization (only their redshift), though our model agrees with this previous interpretation and accurately produces these multiple-images at this redshift. In addition, our model suggests that part of the \emph{radial} arc is also contributed by the left side of the giant arc (system 2). However, it is still plausible that the giant arc consists of only one elongated source. We find that the full giant arc, when projected back to the source plane, corresponds to $\simeq$11 kpc in length, which may indeed be accounted for by a single source. The reproduction of the source is seen in Figure \ref{rep_sys1}.

\begin{figure}
 \begin{center}
 \includegraphics[width=80mm]{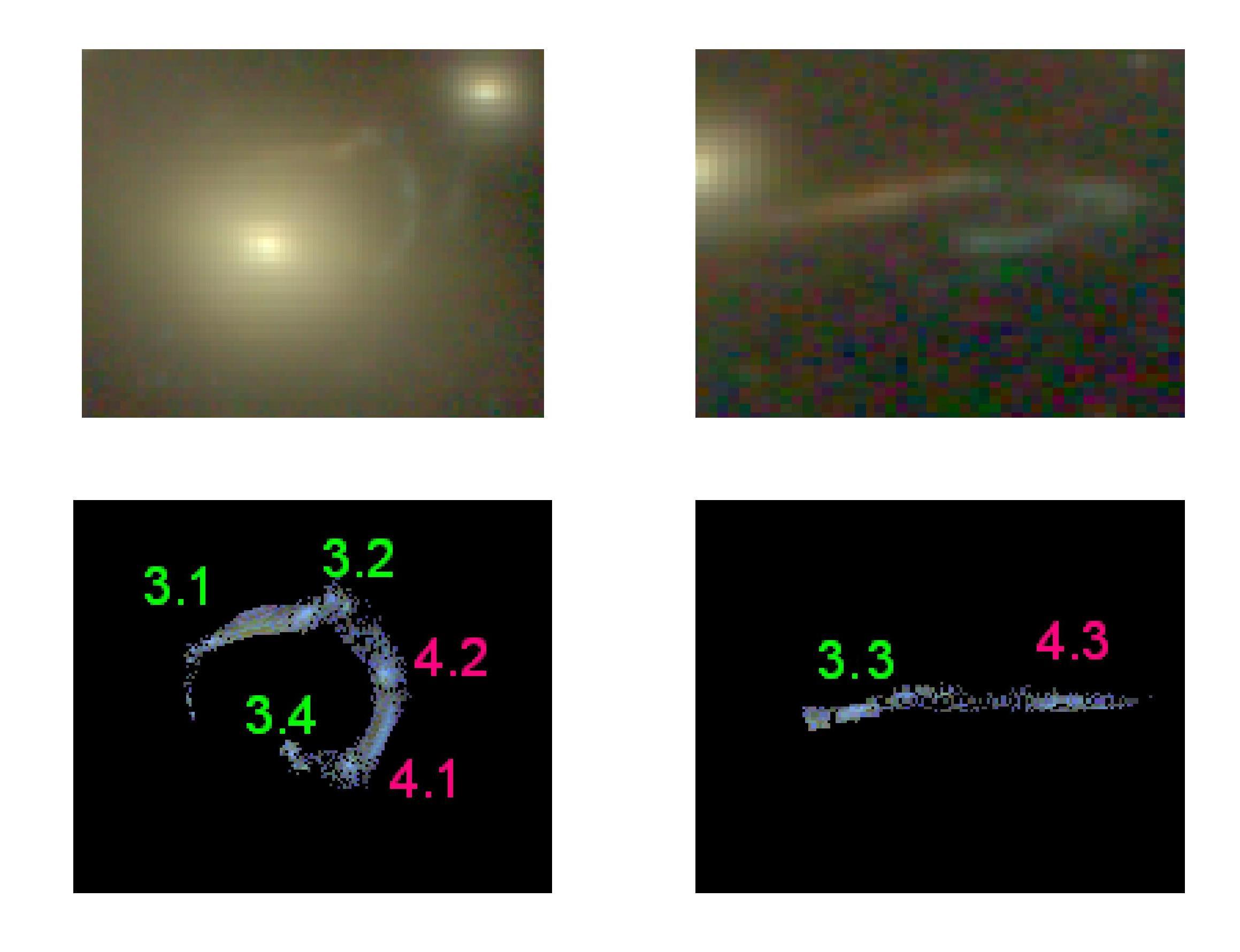}
 \end{center}
\caption{Reproduction of systems 3 and 4 by our model, by delensing jointly
images 3.1 and 4.1 into the source plane with a lensing distance corresponding to $z_s=2.55$, and then relensing the resulting source plane
pixels onto the image plane. Note the different colors of these systems, seen in an WFC3/IR color image.}
\label{rep_sys34}
\end{figure}

$Systems ~3-4:$ Corresponding to a pair of sources, the images
of which are lensed to appear next to each other on the two
sides of a prominent cluster galaxy, as can be seen in Figure
\ref{curves383}. These images were identified by \citet{Smith2001}
who mapped the internal structures in detail, and were
spectroscopically measured by \citet{Newman2011} to be at a redshift
of $z_{s}=2.55$. In addition, our IR/WFC3 images show clearly, for the first time, that
these two systems indeed have two different colors and SEDs, and our
mass model accurately reproduces each as shown in Figure \ref{rep_sys34}. We note that our model suggests that a small part of the radial arc may consist of a counter image of these systems, in addition to the radial images of systems 1 and 2.

Eastwards to image 3.3 there is a faint arc which might be related either to this system or to system 4. This faint extending arc was marked as part of this system by \citet{Smith2001} but omitted in recent analysis \citep{Newman2011}. We find that this faint arc, marked as 3.5/4.4 here, may be related to this system (also yielding a similar photometric redshift of $\sim2.7$), and is reproduced as part of this system by our model if we slightly increase the weight of its neighboring galaxy (RA=02:48:03.42, DEC=-03:32:09.02), see Figure \ref{curves383}. On the other hand, the IR colors do not strongly support connection to systems 3 and 4, and this faint arc might be a locally multiply-lensed separate system. In any case its inclusion has only a negligible and local effect on the mass model.

$System ~5:$ Two images of a multiply-lensed Lyman-break, high redshift galaxy at $z_{s}=6.027$, reported recently by \citet{Richard2011} based on CLASH imaging and Keck spectra. We also identify these two images and measure photometric redshifts of $z \approx 6.01$ and $5.95$.
The high redshift of this system expands substantially the lensing-distance range
thus enabling us to constrain the profile with better accuracy, as discussed in \S \ref{magsec}.

\begin{figure}
 \begin{center}
 \includegraphics[width=90mm]{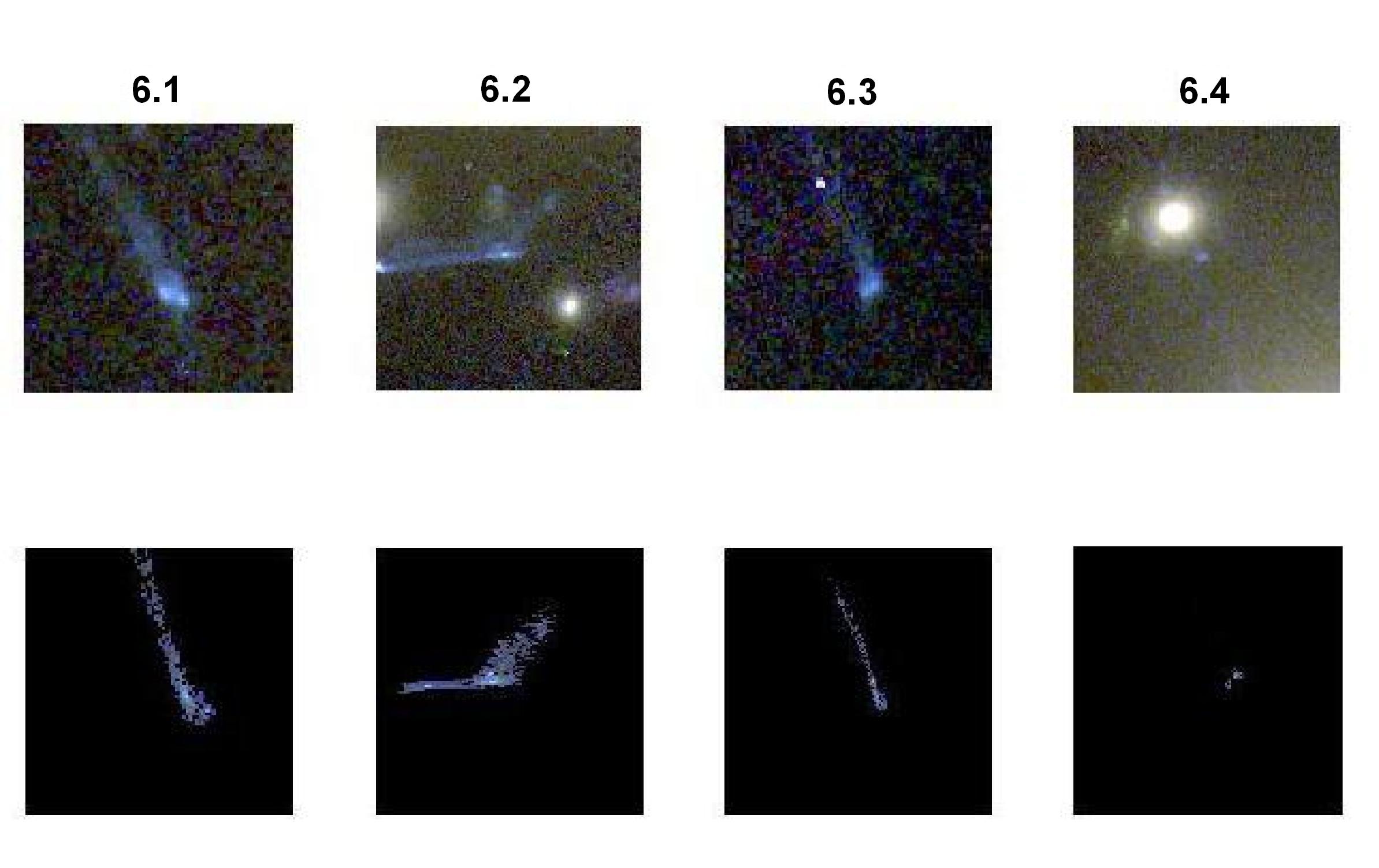}
 \end{center}
\caption{Reproduction of system 6 by our model, by delensing
image 6.2 into the source plane with a lensing distance corresponding to $z_s\simeq2$, and then relensing the resulting source plane
pixels onto the image plane. Our model clearly reproduces accurately the other images in this system.}
\label{rep_sys5}
\end{figure}

\begin{figure}
 \begin{center}
 \includegraphics[width=85mm]{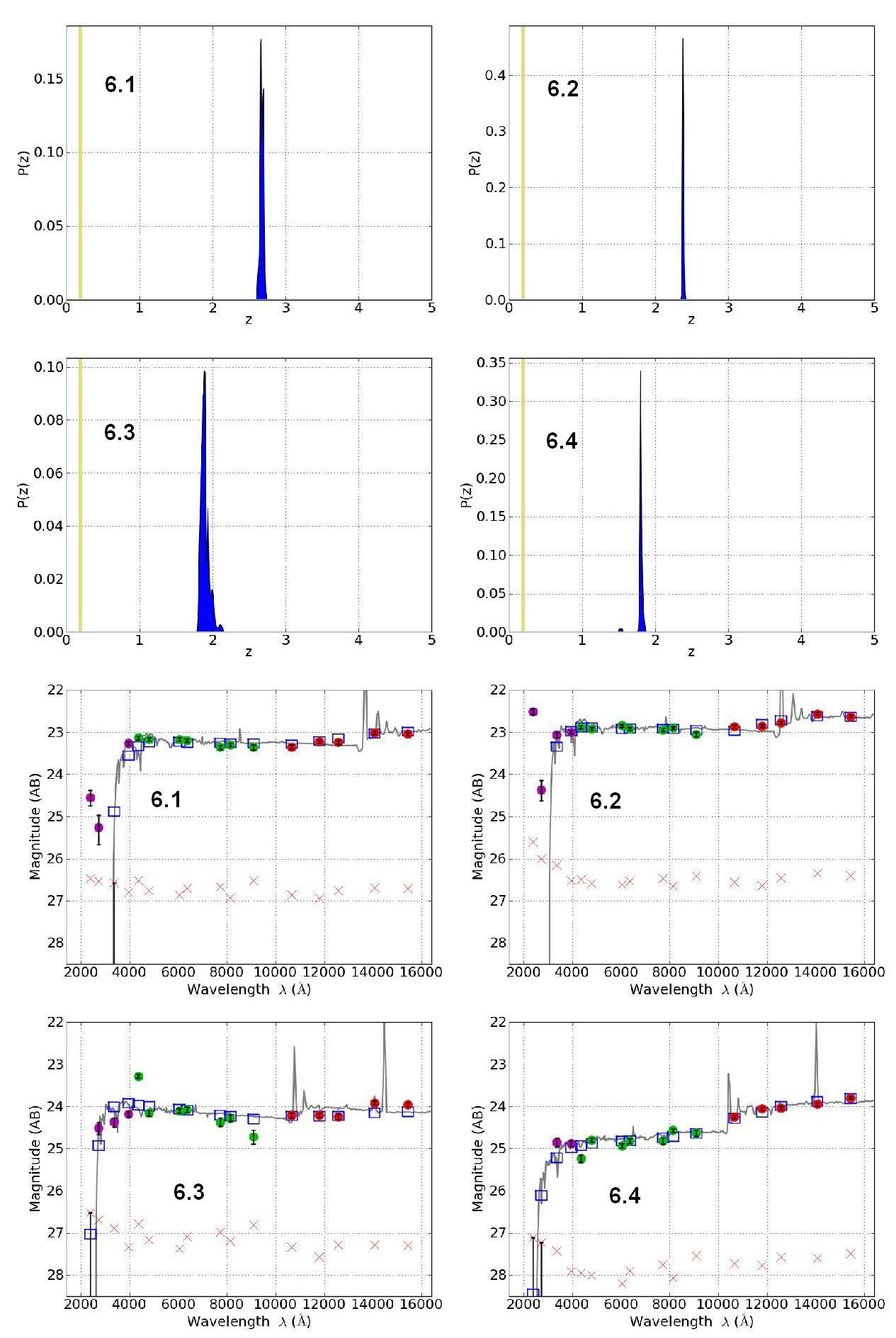}
 \end{center}
\caption{Photo-$z$ distribution (blue) and 16 HST-band SEDs of the four images of system 6 (see Figure \ref{rep_sys5}), generated using BPZ (\S \ref{obs}). As can be seen, the photo-$z$'s support in general our identification of this system, though with some uncertainty and corresponding variation in the SEDs. The yellow stripe in the photo-$z$ distributions corresponds to the cluster redshift at $z=0.189$, and the red crosses in the SED plots mark the $1\sigma$ magnitude detection limit in each filter.}
\label{dist_sys5}
\end{figure}

$System ~6:$ This system consists of 4 blue images with similar
internal details including a brighter white blob, at a typical photometric redshift of $z_{s}\sim2.4$ for this system (see Table \ref{systems}, Figure \ref{dist_sys5}). Our model reproduces these images very well (Figure \ref{rep_sys5}), though it slightly favors a lower redshift of $z_s\simeq2$ but due to the distances involved this is in practice only a $\simeq1\%$ difference in the redshift distance ratio. These images were
matched up for the first time in this work enabled by the deep, high-resolution HST data. Due to the variance in the SEDs and therefore photometric redshifts of the images of this system, we supply also the photo-$z$ distributions and SEDs in Figure \ref{dist_sys5}, so that the reader could more easily assess the plausibility of this system.

\begin{figure}
 \begin{center}
 \includegraphics[width=80mm]{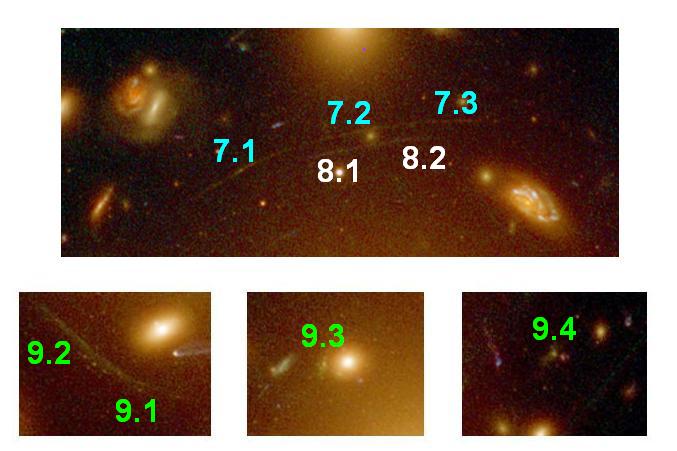}
 \end{center}
\caption{Stamp images of some of our newly identified multiple-images, seen more clearly in this color-composite image constructed from 16 HST bands ranging from the UV to the IR. As can be seen, systems 7 and 8 follow the same symmetry and are clearly multiple images as supported also by our mass model and photo-$z$'s (see Table \ref{systems}). In system 9, images 9.1 and 9.2 are clearly multiply-lensed to form the greenish-looking arc, while 9.3 and 9.4 are the most likely counter images as predicted by our model and coherent photo-$z$'s seen in Figure \ref{dist_sys8}. Other similar looking objects are seen close to these images which might be the actual counter images, but such a degeneracy does not affect the mass model in a noticeable way.}
\label{stamp_sys678}
\end{figure}

$Systems ~7-8:$ Two thin and long arcs following similar symmetry, at a
relatively high redshift of $z\sim4.5$ and $z\sim3$, respectively. Their symmetry especially with regards to the critical curves, shows beyond a doubt that these are multiply-lensed systems (see also Figure \ref{stamp_sys678}), despite being too faint to measure their photometric redshift unambiguously. These images as well were
matched up for the first time in this work.

\begin{figure}
 \begin{center}
 \includegraphics[width=85mm]{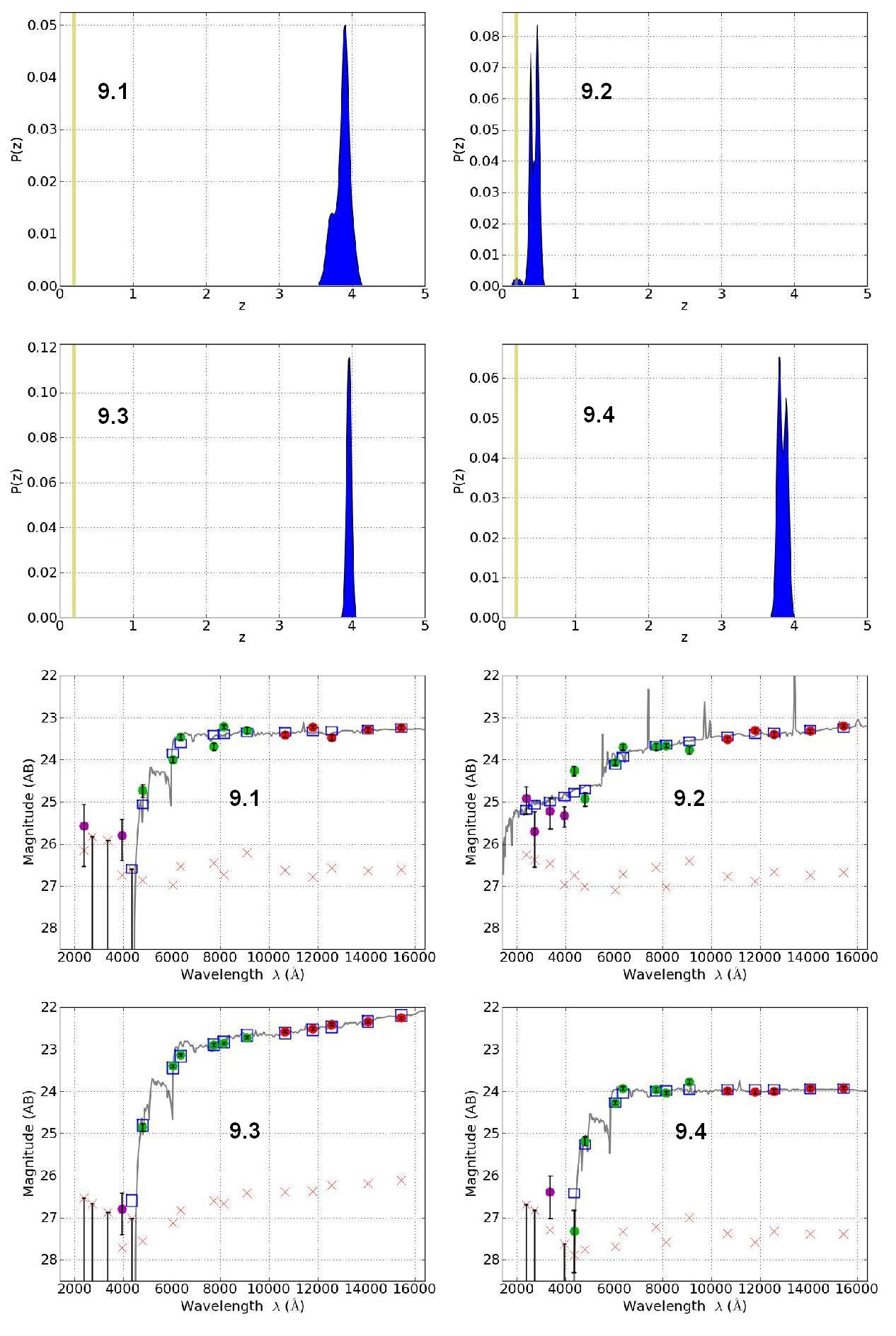}
 \end{center}
\caption{Another example of the photo-$z$ distribution (blue) and 16 HST-band SEDs, generated using BPZ (\S \ref{obs}) and used to match the four images of system 9 (see Figure \ref{curves383}). The photo-$z$ distributions and the SEDs support the identification of this system at $z_{s}\sim3.8$, apart from image 9.2, although note that photo-$z$ analyses of some of its segments do favor a higher-$z$ of $z_{s}\sim3.8$, similar to the other images in this system. The yellow stripe in the photo-$z$ distributions corresponds to the cluster redshift at $z=0.189$, and the red crosses in the SED plots mark the $1\sigma$ magnitude detection limit in each filter.}
\label{dist_sys8}
\end{figure}

$System ~9:$ A faint, wide greenish-looking arc $17\arcsec$ south east of the BCG
(see Figure \ref{curves383}). Our model accurately reproduces this arc
as a double image. In addition, two other small counter-images are predicted, for which we identify the best-matching candidates in the data. These
images were matched up for the first time in this work, and except for image 9.2, show similar photometric redshifts of $\sim3.8$ (see also Figure \ref{dist_sys8}), in agreement with our model prediction. In addition, it should be noted that photo-$z$ analyses of some segments of the arc designated as 9.2 imply indeed a redshift of $\sim3.8$, similar to the other three images of this system. We also acknowledge the possibility that other similar looking objects near-by images 9.3 and 9.4 may be the actual counter images - especially since 9.3 and 9.4 seem slightly brighter than 9.1 and 9.2. Such a degeneracy however does not affect the mass model in a noticeable way.  Due to the variance in the SEDs and therefore photometric redshifts of these images, we supply also their photo-$z$ distributions and SEDs in Figure \ref{dist_sys8}, so that the reader could more easily assess the plausibility of this system.

\subsection{Mass Profile}\label{magsec}

The inner mass profile is accurately constrained by incorporating the
cosmological redshift-distance relation, i.e., the lensing distance of
each system based on the measured spectroscopic or photometric
redshifts. In so doing we normalize our mass model to systems 3 and 4, so
that the normalized scaling factor, $f(d_{ls}/d_{s})$, is equal to 1 for $z_{s}=2.55$.
We then make use of the $z=1.01$ system, and the highest-$z$ system at $z_{s}=6.027$,
in order to expand the $f(d_{ls}/d_{s})$ range, along with the other systems whose photometric or spectroscopic redshifts are incorporated to constrain the profile. The resulting mass profile is seen in Figure \ref{profileAdi}.

\begin{figure}
 \begin{center}
 %\hspace{-2cm}
  %\includegraphics[width=80mm, trim=-10mm 0mm 0mm 0mm,clip]{contours0018.jpg}
   \includegraphics[width=90mm]{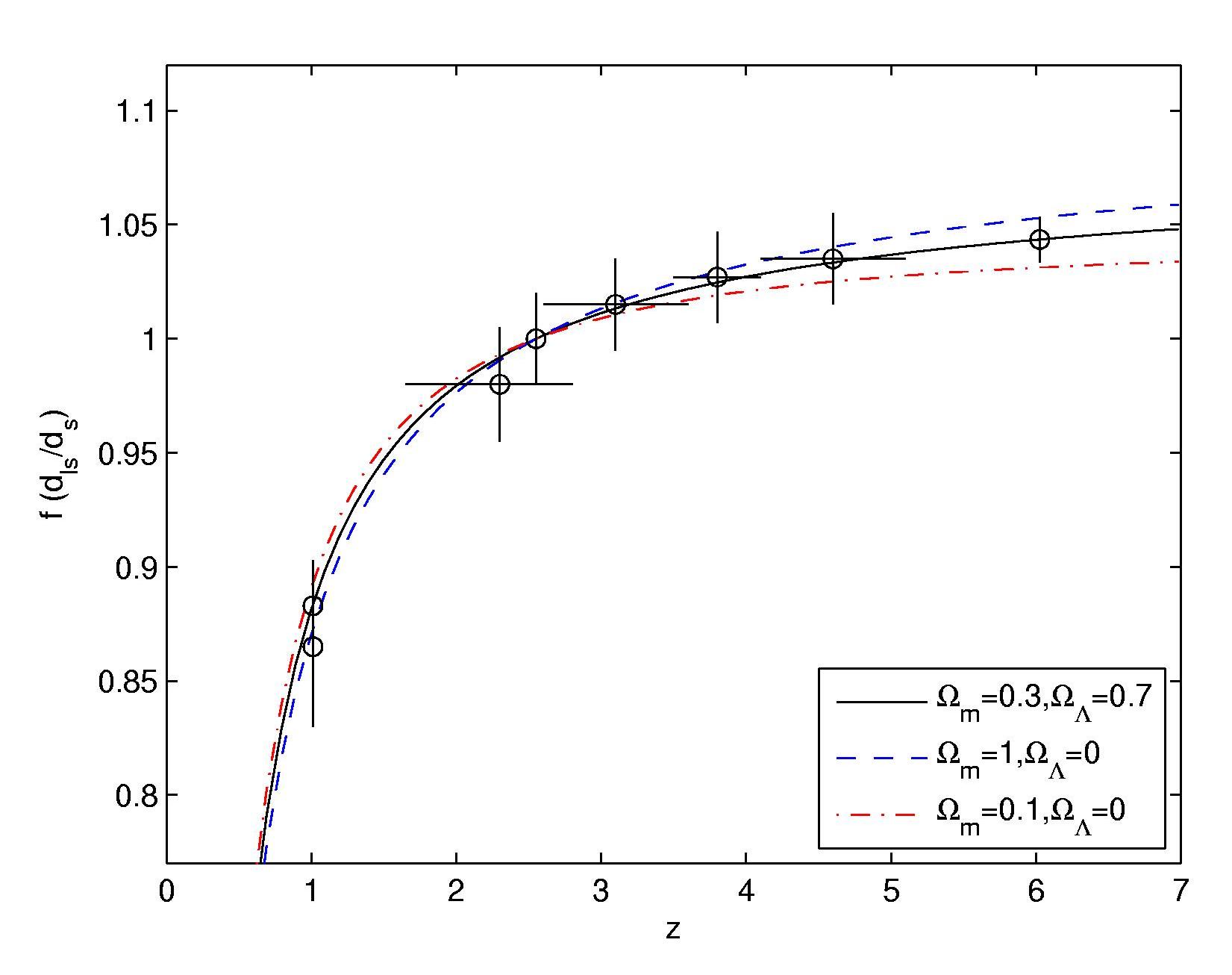}
 \end{center}
\caption{Growth of the scaling factor $f(d_{ls}/d_{s})$ as a function of
redshift, normalized so f=1 at $z=2.55$. Plotted lines are the
expected ratio from the chosen specified cosmological model. The
circles correspond to the multiple-image systems reproduced by our mass model, versus their real spectroscopic or photometric redshift. The
data clearly follow well the relation predicted by the standard
cosmological model. As can be seen, based on only one cluster and due to the low number of multiple images, it may be still hard to discriminate between different cosmologies solely with these data. With a sample of 25 CLASH clusters however, where most clusters are expected to have larger critical area and thus larger numbers of multiple-images spread over a large redshift range, we will be able to put statistically-significant constraints on the cosmological parameters, combining the information from all clusters together.}
\label{dlsds}
\end{figure}

We examine how well the cosmological relation is reproduced by our
model, accounting for all systems with spectroscopic or photometric
redshifts, as shown in Figure \ref{dlsds}. The predicted deflection of the
best fitting model at the redshift of each of these systems clearly lies
along the expected cosmological relation, with a small mean
deviation of only $\Delta_{f}< 0.01$ (see Figure \ref{dlsds}), strengthening the determination of the mass profile slope.

In addition, we note that our mass profile shows consistency with a recent joint lensing, X-ray ,and kinematic analysis by \citet[as read from Fig. 2 therein]{Newman2011}, out to at least twice the Einstein radius where our SL data apply. For example, for the radius of the giant tangential arc (systems 1 and 2), the model of Newman et al. encloses a projected mass of $\simeq2\times10^{13} M_{\odot}$, while our model yields for that radius $\simeq2.2\times10^{13} M_{\odot}$. At higher radii, say a 100 kpc (which is about twice the Einstein radius), both models yield similarly $\simeq6\times10^{13} M_{\odot}$. Due to the different interpretation of the radial arc, some differences are seen in the very inner region, so that for radii of 5-10 kpc ($1.5-3\arcsec$) our model yields $\simeq0.09-0.2\times10^{13} M_{\odot}$, versus $\sim0.05-0.7\times10^{13} M_{\odot}$ for the model by Newman et al.

We combine our SL-based profile with 1D WL distortion and magnification measurements out to and beyond the virial radius ($R_{vir}\simeq 11.3$ arcmin; or $\simeq$2.1 Mpc; corresponding to an overdensity of $\simeq$115 with respect to the critical density of the universe at the cluster redshift), obtained from deep multicolor Subaru imaging (see Figure \ref{profileAdi}). Here we have chosen the BCG position as the center of mass for our mass profile analysis, where our strong-lens modeling shows that the dark-matter center of mass is consistent with the location of the BCG, without any noticeable offset within errors.
The SL profile is obtained in 81 linearly-spaced radial bins from $\theta=2\arcsec$ (excluding the BCG) to $42\arcsec$, including cosmic covariance between radial bins due to the uncorrelated large scale structure, estimated by projecting the nonlinear matter power spectrum out to the median depth of $z_s=2.55$ (see Table 2), following the prescription detailed in \citet{Umetsu2011b}.

The WL mass profile, given in logarithmically-spaced radial bins, was derived using the Bayesian method of \citet{Umetsu2011b,Umetsu2011a} that combines WL tangential-distortion and magnification-bias measurements in a model-independent manner, with the assumption of quasi-circular symmetry in the projected mass distribution.\footnote{This method applies without the axial symmetry approximation in the WL regime where nonlinearity between the surface mass density and observables is negligible.}
The method applies to the full radius range outside the Einstein radius, and is free from the mass-sheet degeneracy,
recovering the absolute mass normalization or equivalently the projected mass $M_{2D}(<\theta_{\rm min}^{\rm WL})$ (corresponding to the first WL bin of Figure 5) interior to the inner radial boundary of WL measurements, $\theta_{\rm min}^{\rm WL}=42\arcsec (\gg r_{E})$. The strong and weak lensing are in excellent agreement where the data overlap, $\theta \simeq 0.7- 1 \arcmin$ ($R \simeq 85-190$ kpc).

For comparison we overplot in Figure \ref{profileAdi} also the recent profile of \citet{Huang2011} derived from the Subaru WL distortion data.
The two profiles are in good agreement and very similar in the WL regime, but the Huang et al. profile is slightly underestimated in the inner region relative to our SL data. Our secure background selection method \citep{Medezinski2010, Medezinski2011} carefully combines all color and clustering information to identify blue and red background galaxies in color-color space ($B-R_{C}$ vs. $R_{C}-z'$), minimizing contamination by unlensed cluster and foreground galaxies. It is important to stress that combining independent weak and strong lensing allows us to recover the full radial profile and ensure internal consistency in the region of overlap.

We consider a generalized parametrization of the NFW \citep{Navarro1996} model of the
following form \citep{Zhao1996, JingSuto2000}:
\begin{eqnarray}
\label{eq:NFW}
 \rho_{\rm NFW}(r)= \frac{\rho_s}{(r/r_s)^{\alpha} (1+r/r_s)^{3-\alpha}}~~,
\end{eqnarray}
where $\rho_s$ is the characteristic density, $r_s$ is the characteristic scale radius, and $\alpha$ is the inner slope of the density profile.  This model has an asymptotic outer slope of $\gamma_{3D}(r)\equiv d\ln{\rho}/d\ln{r}=-3$ $(r\to \infty)$,
and reduces to the NFW model for $\alpha=1$.

We refer to the profile
given by equation \ref{eq:NFW} as the generalized NFW (gNFW,
hereafter) profile. It is useful to introduce the radius $r_{-2}$
at which the logarithmic slope of the density is isothermal,
i.e., $\gamma_{3D} = -2$. For the gNFW profile, $r_{-2} = (2 - \alpha)r_{s}$, and
thus the corresponding concentration parameter reduces to
$c_{-2}\equiv r_{vir}/r_{-2} = c_{vir}/(2-\alpha)$. We specify the gNFW model with
the central cusp slope, $\alpha$, the halo virial mass, $M_{vir}$, and the
concentration, $c_{-2} = c_{vir}/(2-\alpha)$.

The joint SL+WL NFW fit yields $M_{vir}=(5.37^{+0.70}_{-0.63}\pm 0.26) \times 10^{14}M_{\odot}/h$ (or $M_{vir}\simeq 7.67 \times 10^{14}M_{\odot}$) and a concentration parameter of $c_{vir} = 8.77^{+0.44}_{-0.42} \pm 0.23$, with a minimized $\chi^2$ ($\chi^2_{\rm min}$) value of 78.7/90 with respect to the degrees of freedom (dof), corresponding to a goodness-of-fit of $Q=0.798$. Note that the values quoted include the statistical followed by the systematic uncertainty at a 68\% confidence level.
The systematic errors were estimated by changing the outer radial boundary of SL bins from $\theta=42\arcsec (=\theta_{\rm min}^{\rm WL})$ to $2 r_{E} (\simeq 33\arcsec)$.

Only a very slight improvement in the fit is obtained by implementing the gNFW form described in eq. \ref{eq:NFW}. A joint SL+WL gNFW fit yields $M_{vir}=(5.94^{+1.05}_{-0.87}\pm 0.71) \times 10^{14}M_{\odot}/h$ (or $M_{vir}\simeq 8.49 \times 10^{14}M_{\odot}$), a concentration parameter of $c_{-2} = 7.95^{+0.89}_{-0.90}\pm 0.86$, and $\alpha=1.078^{+0.069}_{-0.073} \pm 0.059$, with $\chi^2_{\rm min}/{\rm dof}=77.5/89$ and $Q=0.802$. The central cusp slope $\alpha$ is consistent with unity, so that the overall fit is similar to a simple NFW, as also shown by the quoted $\chi^{2}$ and $Q$ values. These results are consistent with the values quoted by \citet{Huang2011} ($M_{vir}=5.28^{+1.86}_{-1.34}\times 10^{14}M_\odot/h, c_{vir}=5.68^{+2.11}_{-1.60}$), but the concentration is higher for example than the Chandra X-ray based gNFW fit by \citet{SchmidtAllen2007}.

\begin{figure}
 \begin{center}
   \includegraphics[width=85mm]{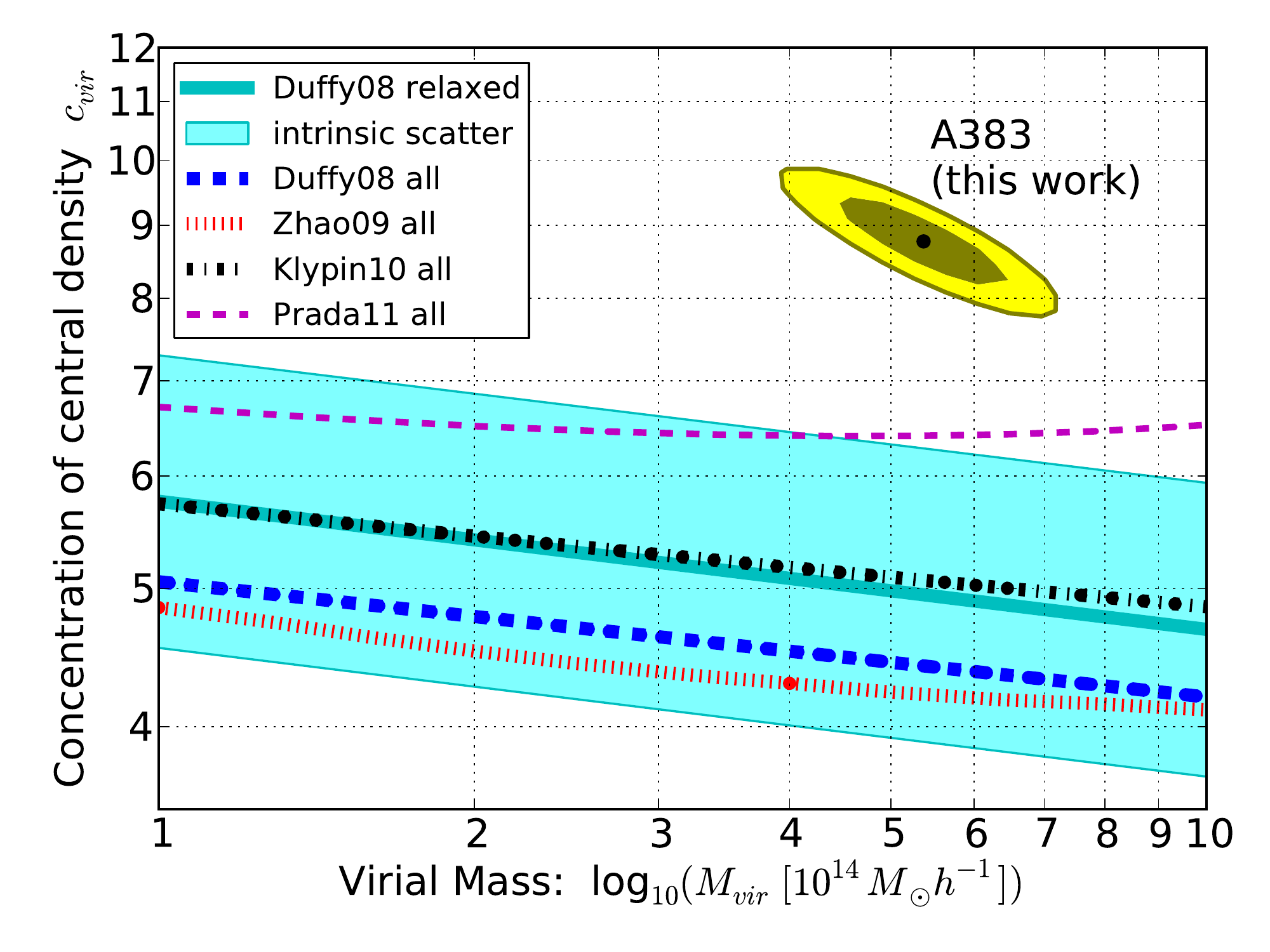}
 \end{center}
\caption{The joint SL+WL NFW fit of A383 (data point with $1\sigma$ and $2\sigma$ confidence level contours) presented on the $c$--$M$ plane, compared to expectations from simulations. Overplotted are the expected $c$--$M$ relations and their 1$\sigma$ uncertainties, presented in \citet{Duffy2008} for their relaxed sample, scaled to $z_c=0.189$ (blue band). Also plotted are $c$--$M$ relations for the full-sample clusters from \citet{Duffy2008, Zhao2009, Klypin2010,Prada2011SimCDM}. As has been found for other observed clusters (e.g., \citealt{Broadhurst2008, Oguri2009, Zitrin2010}) A383 appears to have a higher concentration than simulated clusters with similar masses and redshifts. We note that the WL data used here are based on 1D analysis, while the concentration should be influenced by the triaxiality, and this result will be revised in our following WL paper using 2D analysis (Umetsu et al., in preparation).}
\label{CM}
\end{figure}

We find that A383 lies above the standard $c$--$M$
relation (Figure \ref{CM}), similar to several other
well-known clusters for which detailed lensing-based mass profiles have
been constructed, adding to the claimed tension
with the standard $\Lambda$CDM model (\citealt{Broadhurst2008, Umetsu2010, Umetsu2011a, Zitrin2010}; see also \citealt{SadehRephaeli2008}). Still, the overall level of systematic uncertainties may be
too large to allow a definite conclusion regarding a clear inconsistency with
$\Lambda$CDM predictions based on only a handful of clusters. This, in fact, is one of the primary goals of our CLASH program. Moreover, the WL data used here are based on 1D analysis, while the concentration should be influenced by the triaxiality or other line-of-sight background structures, and this result will be revised in our following WL papers, using 2D analysis (Umetsu et al., in preparation) and a joint SL+WL non-parametric reconstruction method (Merten et al., in preparation).

In addition, we note that several dips are seen in our WL tangential distortion data in outer radii, resulting in positive perturbations in the $\kappa$ profile. We confirmed, by visual inspection in deep color Subaru imaging, that these correspond to several higher-redshift background structures near the field of A383 (see also \citealt{Okabe2010}). In fact, the field is quite rich in such background structures, some of which are slightly magnified by the A383 foreground lens, as we will elaborate in our upcoming papers devoted to this configuration (Umetsu et al., Zitrin et al., in preparation).

\subsection{Brightest Cluster Galaxy}\label{bcgsec}

Due to the presence of the radial arc in its halo, we may also constrain the mass enclosed within the BCG, and the corresponding M/L ratio. We find that the BCG encloses a \emph{projected} mass of $1.14\pm0.3 \times
10^{12} M_{\odot}$ within a radius of $\simeq6\arcsec$ ($\simeq$19 kpc) after
subtracting the interpolated smooth DM component ($\simeq4.1\times
10^{12} M_{\odot}$ inside this aperture). For comparison, recent stellar velocity-dispersion measurements of the BCG in A383 yield $\sigma\simeq450\pm40$ at this radius (as read from Figure 2 in \citealt{Newman2011}), which translates to a projected mass of $0.89^{+0.22}_{-0.15} \times10^{12} M_{\odot}$, in agreement with our result.

We measure the BCG flux in several optical ACS bands, to obtain an average B-band
luminosity of $\simeq8\pm0.3 \times 10^{10} L_{\odot}$, within the aperture of $\simeq6\arcsec$ ($\simeq$19 kpc; fluxes were converted to luminosities using the LRG template described in \citealt{Benitez2009BAU_LRG}). This yields a typical $M/L_B$ of
$\sim14~(M/L)_{\odot}$ in this region, similar to other lensing based BCG masses in well-studied clusters (e.g., \citealt{Gavazzi2003} for MS 2137-2353, \citealt{ZitrinBroadhurst2009} for MACS J1149.5+2223, \citealt{Limousin2008, Zitrin2010} for Abell 1703), though of course the degeneracy between the BCG DM halo and the overall subtracted smooth cluster halo is still unknown.

\subsection{Modeling Accuracy and Uncertainty}

In general, since the deflection angle depends on the distance-redshift ratio ($D_{ls}/D_{s}$), the SL modeling uncertainty, particularly with regards to the mass profile, is primarily coupled to the redshift measurement accuracy of the multiple-systems. In A383, five systems at three different redshifts have spectroscopic measurements, while the four other systems found in this work importantly supply four more constraints on the mass profile. Our 16-band ACS/WFC3 imaging, allows us to derive robust photo-$z$'s for all multiply-lensed systems discussed in this work, which we verify by using both the BPZ method, and the LPZ method (\S \ref{obs}).

Still, due to the low number of parameters in our modeling, which constitutes a huge advantage for finding multiple-images and producing efficiently well-constrained mass distributions, some local inaccuracy can be expected. Our best fit model for A383 reproduces all multiple-images described in this work within $\simeq2.5\arcsec$ from their real location given their measured redshifts, aside from one image candidate belonging to system 9 (see \S \ref{Mimages}) which is reproduced $\sim4\arcsec$ from its real location. In addition, we note that in our best-fit model, constrained by all systems together, there is a slight offset of $\simeq1\arcsec$ in the reproduced location of the radial arc, implying a slight inaccuracy in the BCG's very inner mass profile in that model. Note however that this does not affect the result in \S \ref{bcgsec}, which was verified by complementary models in which the radial arc is very accurately reproduced (but the fit is overall somewhat poorer taking into account all other systems), thus reliably constraining the mass enclosed within the corresponding radius.

The average image-plane reproduction uncertainty of our best-fit model is $1.68\arcsec$ per image in total, with an image-plane $rms$ of $1.95\arcsec$ including all 27 multiply-lensed images. This image-plane $rms$ is, for example, higher than that reported recently by \citet{Newman2011} for A383 ($rms=0.27\arcsec$) based on only four systems in two different spectroscopic redshifts, but is typical to most parametric-method reconstructions, when many multiple-systems are present. For example, \citet{Broadhurst2005a} achieved an $rms$ of $3.2\arcsec$ per image for
Abell 1689, and later \citet{Halkola2006} reported an $rms$ of $2.7\arcsec$ per image for that cluster, while \citet{Zitrin2009b} produced an $rms$ of $\simeq2.5\arcsec$ for Cl0024. These values are comparable with our current model $rms$, taking into account the difference in the critical area.

In general, a higher number of parameters would supply a more accurate solution, however the efficiency of a
model and the confidence in it decrease substantially as more parameters are
added to the minimization procedure, especially if these are arbitrary non-physical parameters as may be the case in other (non-parametric) methods.
We have shown here as well as in many previous examples (see also \S \ref{model}) that our method, with a minimum number of free parameters, built on simple physical considerations (see \citealt{Zitrin2009b} for full details), does a very good job in finding new multiply-lensed systems, and thus in constraining the deflection field, and accordingly, the mass distribution and profile.

\section{Comparison to Numerical Simulations}\label{simu}

We now compare our derived A383 mass distribution with cluster halos obtained from hydrodynamical simulations in the framework of the $\Lambda$CDM cosmology. The analysis we make here is inspired by the work of \citet{Meneghetti2011}, where the Einstein ring sizes and the lensing cross sections of 12 massive MACS clusters modeled by \citet{Zitrin2011a} were compared with those expected from similar halos in the {\sc MareNostrum Universe} cosmological simulation \citep{Gottlober2007, Meneghetti2010a, Fedeli2010}. This is a $500^{3}\;h^{-3}$Mpc$^3$ volume filled with $1024^3$ DM and $1024^3$ gas particles, evolved in the framework of a cosmological model with $\Omega_m=0.3$, $\Omega_\Lambda=0.7$, and $\sigma_8=0.9$. More details about this simulation can be found in \citet{Gottlober2007}. For our comparison, we use halos extracted from the same cosmological box, for which the median Einstein ring sizes and the cross sections for giant arcs (defined as having length-to-width ratios larger than 7.5), were readily computed as in \citet{Meneghetti2011}. To account for the lack of star formation in these simulations, we added to each halo a component mimicking the presence of a massive galaxy at the cluster center, following the method employed in \citet{Meneghetti2003}. The galaxy was modeled with pseudo-isothermal mass distribution (see e.g. \citealt{Donnarumma2011}) with a velocity dispersion of $\sigma=300$ km/s and a cut-off radius of $23$ kpc.

We use the deflection angle maps of the A383 SL model presented here, and use them to perform a ray-tracing simulation. We stress that such a simulation is completely consistent with those performed for each simulated halo. A large number of artificial elliptical sources is used to populate the source plane at $z_s=2$, which are distributed on adaptive grids with increasing spatial resolution towards the caustics, in order to sample with greater accuracy the regions where sources are strongly magnified. By counting the sources that are lensed as giant arcs, we measure the lensing cross section, which is the area surrounding the caustics where sources must be located in order to produce images with length-to-width ratios larger than 7.5.
The deflection angle maps also allow to measure the cluster median Einstein ring, defined as the median distance of the critical points from the cluster center. In the following discussion, we refer to the Einstein radius for sources at redshift $z_s=2$. Note also that the median Einstein radius is defined differently than the simple effective radius of the critical area which is usually used and was implemented throughout this work in order to compare to other results. In this section only, we use the $median$ Einstein radius, in order to be consistent with previous work based on these simulations and since it usually better correlates with the lensing cross-section (see \citealt{Meneghetti2011}).

\begin{figure}
 \begin{center}
 %\hspace{-2cm}
  %\includegraphics[width=80mm, trim=-10mm 0mm 0mm 0mm,clip]{contours0018.jpg}
   \includegraphics[width=80mm]{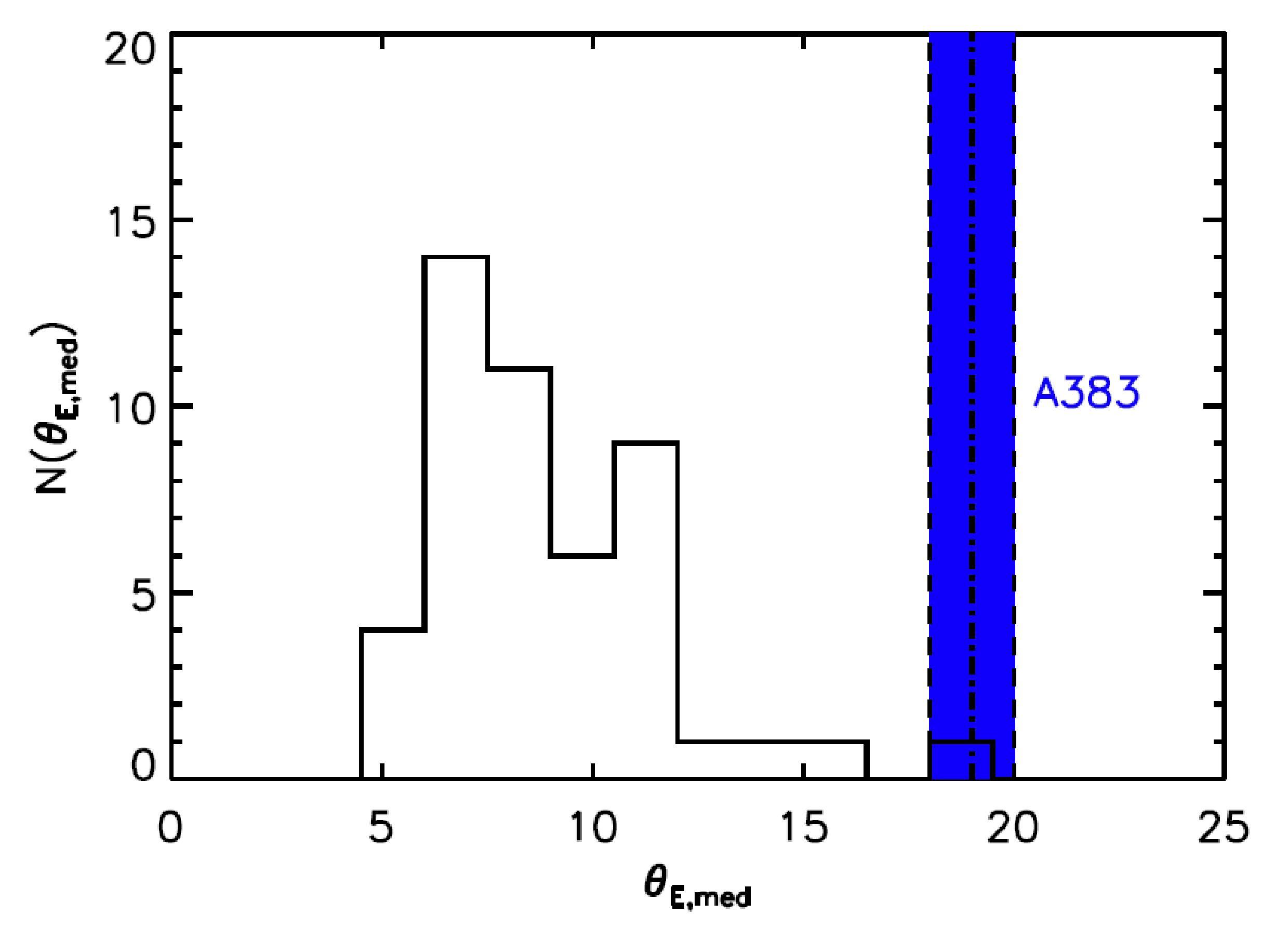}
 \end{center}
\caption{Distribution of the $median$ Einstein radii of halos with $6\times10^{14}\;h^{-1}M_\odot \leq M_{vir} \leq 7\times10^{14}\;h^{-1}M_\odot$ extracted from the {\sc MareNostrum Universe} snapshot at $z\sim0.19$. The blue-shaded region shows the size of the median Einstein radius of A383 with its error-bar. See \S \ref{simu} for more details and the definition of the $median$ Einstein radius.}
\label{thetaE}
\end{figure}

By doing this analysis, we find that A383 has a median Einstein radius  $\theta_{\rm med}=19.0\pm1 \arcsec$ , and a lensing cross section $\sigma=2.49\pm0.9\times 10^{-3}\;h^{-2}$Mpc$^2$. Comparing these results with the distributions of $\theta_{\rm med}$ and $\sigma$ of halos with similar mass in the {\sc MareNostrum Universe}, we find that A383 is a remarkably strong lens, given its relatively small mass.  Indeed, the majority of simulated clusters with $6\times 10^{14} \leq M_{vir} \leq 7\times 10^{14} \;h^{-1}M_\odot$ have much smaller critical curves and cross sections. For example, the mean Einstein radius and lensing cross section of such sample are $8.9 \arcsec$ and $5.6\times 10^{-4}\;h^{-2}$Mpc$^2$, respectively. The values measured for A383 exceed the maximal values measured in the simulations, which are $18.14\arcsec$  and $2.26\times 10^{-3}\;h^{-2}$Mpc$^2$, respectively. As an example, the distribution of median Einstein radii for simulated clusters is shown in Fig.\ref{thetaE}. As shown by \citet{Meneghetti2011}, the lensing cross-section is tightly correlated to the median Einstein radius. We note also that each cluster in the {\sc MareNostrum Universe} was projected along three independent lines-of-sight in order to account for possible projection effects, i.e. for including cases where clusters are seen nearly along their major axis. The largest Einstein radii and cross sections are indeed produced by clusters whose major axis is almost perfectly aligned to the line-of-sight. Nevertheless, it is difficult to find a cluster that matches the mass and the strong lensing efficiency of A383 among the simulated halos. Extending the upper mass limit of the simulated sample to $8\times 10^{14}\;h^{-1}M_\odot$, we find that only two cluster projections have Einstein radii and lensing cross sections larger than those measured for A383, which is still in the $>99\%$ tail of the corresponding distributions.

\citet{Meneghetti2010b} showed that the concentrations estimated from the projected mass distribution of strong lensing clusters are on-average biased high, i.e. they are higher than the corresponding concentrations measured from the three-dimensional cluster mass distributions (see also \citealt{Hennawi2007}). Such bias depends on the cluster mass, redshift, and lensing cross section. For a fixed mass and redshift, clusters with large lensing cross sections are typically affected by a more severe concentration bias. As done by \citet{Meneghetti2011}, we use the {\sc MareNostrum Universe} clusters to estimate a lower limit of the concentration bias for A383-like clusters. To do that, we select the numerically simulated halos with redshift and mass matching those of A383 and lensing cross section $\sigma > 10^{-4}\;h^{-2}$Mpc$^2$. This limit was set in order to have a statistically significant sample of simulated halos. The median ratio of $c_{2D}/c_{3D}$ for these lenses is $\sim 1.35$. Thus, for objects with a lensing efficiency as high as in A383, the concentration measured from lensing is expected to be $>35\%$ higher than their true 3D-concentration. This expectation agrees well with a recent work by \citet{Morandi2011On383} estimating the triaxial shape of A383. \citet{Morandi2011On383} deduced by a joint analysis of X-ray and SL measurements (which are commensurate with our analysis), a concentration of $c_{vir} \sim6.1$, while we obtained indeed a $44\%$ higher value in our 2D analysis, $c_{vir} \simeq8.8$ (see \S \ref{magsec}).

\section{Summary}

In this work we have presented a new detailed lensing analysis of the
galaxy cluster A383 in multi-band ACS/WFC3 images. Our
well-established modeling method \citep{Broadhurst2005a, Zitrin2009b, Zitrin2010, Merten2011, Zitrin2011a, Zitrin2011c, Zitrin2011b} has identified 13 multiply-lensed images and candidates, so that in total 27 images of 9
different sources were incorporated to fully constrain the fit. Though more
lensed candidates may generally be found in this lensing field with further
careful effort, the resulting model is clearly fully
constrained by these multiple systems.

The accurate photometric redshifts of the newly
found multiple-systems enabled by the extensive multi-band HST imaging allow for
the most secure lensing-based determination of the inner mass profile of A383
to date, through the cosmological lensing-distance ratio, and imply a
mass profile of $d\log \Sigma/d\log r\simeq -0.6\pm 0.1$, similar to other well-known relaxed clusters, and in excellent agreement with WL
analysis from wide-field Subaru data (Umetsu et al., in preparation, see also Figure \ref{profileAdi}). In addition, we note that our mass profile is generally consistent with a recent joint lensing, X-ray ,and kinematic analysis by \citet{Newman2011}, out to at least twice the Einstein radius where our SL data apply.

In Figure \ref{curves383} we plotted the critical curves along with
the multiple images found and used in this work. For a source at
$z_s=2.55$, the effective Einstein radius $r_{E}=16.3\pm2\arcsec$, or $\simeq$52 kpc at the redshift of
the cluster. This critical curve encloses a projected mass of $M= 2.4\pm0.2 \times
10^{13}M_{\odot}$, in agreement with other
published results (e.g., \citealt{Smith2001, Newman2011}).

We compared the properties of A383 with clusters of similar mass drawn from the {\sc MareNostrum Universe} numerical simulation (see \S \ref{simu}). We find that A383 is a remarkably strong lens, given its relatively small mass. The majority of simulated clusters $6\times 10^{14} \leq M_{vir} \leq 7\times 10^{14} \;h^{-1}M_\odot$ have much smaller critical curves and lensing cross sections. The largest Einstein radii and cross sections are produced by clusters whose major axis is almost perfectly aligned to the line-of-sight. Even with this taken into account, it is difficult to find a cluster that matches the mass and the strong lensing efficiency of A383 among the simulated halos, so that A383 lies at the $>99\%$ tail of the corresponding distributions (Figure \ref{thetaE}). Accordingly, for objects with a lensing efficiency as high as in A383, the concentration measured from lensing is expected to be $>35\%$ higher than their true 3D-concentration, in agreement with recent results \citep[e.g.,][]{Morandi2011On383}.

A383 is the first cluster observed and analyzed in the CLASH
framework (see \S1). As we have shown, despite the relatively small
Einstein radius and correspondingly low number of multiply-lensed images,
the remarkable 16 filter imaging allowed us to immediately uncover several new multiple-systems. With a statistical sample of 25 massive galaxy clusters being deeply
imaged with HST, we should be able to test structure formation models with unprecedented precision.

%The CLASH observations, when combined with complementary wide-field optical and
%X-ray imaging, is expected to constitute a leap in the quality and
%quantity of SL data, enabling us to measure the DM profile shapes and
%mass concentrations from hundreds of multiply-imaged sources,
%providing precise (up to $\sim10\%$) observational challenges to scenarios for
%the DM mass distribution, discover distant high-z galaxies, and detect
%new supernovae, thus shedding more light on the dark Universe and the
%cosmological parameters.

\section*{acknowledgments}

We thank the anonymous referee of this paper for useful comments that improved the manuscript. The CLASH Multi-Cycle Treasury Program (GO-12065) is based on observations made with the NASA/ESA Hubble Space Telescope. We are especially grateful to our program coordinator Beth Perrillo for her expert assistance in implementing the HST observations in this program. We thank Jay Anderson and Norman Grogin for providing the ACS CTE and bias striping correction algorithms used in our data pipeline. We are grateful to Stefan Gottl\"ober and Gustavo Yepes for giving us access to the MareNostrum Universe simulation and to Stefano Ettori for helpful discussions. This research is supported in part by NASA grant HSTGO12065.01-A, the Israel Science Foundation, the Baden-W\"uerttemberg Foundation, the German Science Foundation (Transregio TR 33), Spanish MICINN grant YA2010-22111-C03-00, funding from the Junta de Andaluc\'ia Proyecto de Excelencia NBL2003, INAF contracts ASI-INAF I/009/10/0, ASI-INAF I/023/05/0, ASI-INAF I/088/06/0, PRIN INAF 2009, and PRIN INAF 2010, NSF CAREER grant AST-0847157, the UK's STFC, the Royal Society, the Wolfson Foundation, the DFG cluster of excellence Origin and Structure of
the Universe, and National Science Council of Taiwan grant NSC97-2112-M-001-020-MY3. Part of this work is based on data collected at the Subaru Telescope, which is operated by the National Astronomical Society of Japan. AZ acknowledges support by the John Bahcall excellence prize. The HST science operations center, the Space Telescope Science Institute, is operated by the Association of Universities for Research in Astronomy, Inc. under NASA contract NAS 5-26555.

\bibliographystyle{apj}
\bibliography{outNew383}

\begin{thebibliography}{90}
\expandafter\ifx\csname natexlab\endcsname\relax\def\natexlab#1{#1}\fi

\bibitem[{{Anderson} \& {Bedin}(2010)}]{AndersonBedin2010}
{Anderson}, J., \& {Bedin}, L.~R. 2010, \pasp, 122, 1035

\bibitem[{{Arnouts} {et~al.}(1999){Arnouts}, {Cristiani}, {Moscardini},
  {Matarrese}, {Lucchin}, {Fontana}, \& {Giallongo}}]{ArnoutsLPZ1999}
{Arnouts}, S., {Cristiani}, S., {Moscardini}, L., {Matarrese}, S., {Lucchin},
  F., {Fontana}, A., \& {Giallongo}, E. 1999, \mnras, 310, 540

\bibitem[{{Ben{\'{\i}}tez}(2000)}]{Benitez2000}
{Ben{\'{\i}}tez}, N. 2000, \apj, 536, 571

\bibitem[{{Ben{\'{\i}}tez} {et~al.}(2004){Ben{\'{\i}}tez}, {Ford}, {Bouwens},
  {Menanteau}, {Blakeslee}, {Gronwall}, {Illingworth}, {Meurer}, {Broadhurst},
  {Clampin}, {et~al.}}]{Benitez2004}
{Ben{\'{\i}}tez}, N., {et~al.} 2004, \apjs, 150, 1

\bibitem[{{Ben{\'{\i}}tez} {et~al.}(2009{\natexlab{a}}){Ben{\'{\i}}tez},
  {Gazta{\~n}aga}, {Miquel}, {Castander}, {Moles}, {Crocce},
  {Fern{\'a}ndez-Soto}, {Fosalba}, {Ballesteros}, {Campa},
  {et~al.}}]{Benitez2009BAU_LRG}
---. 2009{\natexlab{a}}, \apj, 691, 241

\bibitem[{{Ben{\'{\i}}tez} {et~al.}(2009{\natexlab{b}}){Ben{\'{\i}}tez},
  {Moles}, {Aguerri}, {Alfaro}, {Broadhurst}, {Cabrera-Ca{\~n}o}, {Castander},
  {Cepa}, {Cervi{\~n}o}, {Crist{\'o}bal-Hornillos}, {Fern{\'a}ndez-Soto},
  {Gonz{\'a}lez Delgado}, {et~al.}}]{Benitez2009PHOTZ}
---. 2009{\natexlab{b}}, \apjl, 692, L5

\bibitem[{{Broadhurst} {et~al.}(2005{\natexlab{a}}){Broadhurst}, {Takada},
  {Umetsu}, {Kong}, {Arimoto}, {Chiba}, \& {Futamase}}]{Broadhurst2005b}
{Broadhurst}, T., {Takada}, M., {Umetsu}, K., {Kong}, X., {Arimoto}, N.,
  {Chiba}, M., \& {Futamase}, T. 2005{\natexlab{a}}, \apjl, 619, L143

\bibitem[{{Broadhurst} {et~al.}(2008){Broadhurst}, {Umetsu}, {Medezinski},
  {Oguri}, \& {Rephaeli}}]{Broadhurst2008}
{Broadhurst}, T., {Umetsu}, K., {Medezinski}, E., {Oguri}, M., \& {Rephaeli},
  Y. 2008, \apjl, 685, L9

\bibitem[{{Broadhurst} {et~al.}(2005{\natexlab{b}}){Broadhurst},
  {Ben{\'{\i}}tez}, {Coe}, {Sharon}, {Zekser}, {White}, {Ford}, {Bouwens},
  {Blakeslee}, {Clampin}, {et~al.}}]{Broadhurst2005a}
{Broadhurst}, T., {et~al.} 2005{\natexlab{b}}, \apj, 621, 53

\bibitem[{{Bruzual} \& {Charlot}(2003)}]{BC03}
{Bruzual}, G., \& {Charlot}, S. 2003, \mnras, 344, 1000

\bibitem[{{Bullock} {et~al.}(2001){Bullock}, {Kolatt}, {Sigad}, {Somerville},
  {Kravtsov}, {Klypin}, {Primack}, \& {Dekel}}]{Bullock2001}
{Bullock}, J.~S., {Kolatt}, T.~S., {Sigad}, Y., {Somerville}, R.~S.,
  {Kravtsov}, A.~V., {Klypin}, A.~A., {Primack}, J.~R., \& {Dekel}, A. 2001,
  \mnras, 321, 559

\bibitem[{{Coe} {et~al.}(2010){Coe}, {Ben{\'{\i}}tez}, {Broadhurst}, \&
  {Moustakas}}]{Coe2010}
{Coe}, D., {Ben{\'{\i}}tez}, N., {Broadhurst}, T., \& {Moustakas}, L.~A. 2010,
  \apj, 723, 1678

\bibitem[{{Coe} {et~al.}(2006){Coe}, {Ben{\'{\i}}tez}, {S{\'a}nchez}, {Jee},
  {Bouwens}, \& {Ford}}]{Coe2006}
{Coe}, D., {Ben{\'{\i}}tez}, N., {S{\'a}nchez}, S.~F., {Jee}, M., {Bouwens},
  R., \& {Ford}, H. 2006, \aj, 132, 926

\bibitem[{{Collins} {et~al.}(2009){Collins}, {Stott}, {Hilton}, {Kay},
  {Stanford}, {Davidson}, {Hosmer}, {Hoyle}, {Liddle}, {Lloyd-Davies},
  {et~al.}}]{Collins2009}
{Collins}, C.~A., {et~al.} 2009, \nat, 458, 603

\bibitem[{{Corless} \& {King}(2009)}]{CorlessKing2009}
{Corless}, V.~L., \& {King}, L.~J. 2009, \mnras, 396, 315

\bibitem[{{Daddi} {et~al.}(2007){Daddi}, {Dickinson}, {Morrison}, {Chary},
  {Cimatti}, {Elbaz}, {Frayer}, {Renzini}, {Pope}, {Alexander},
  {et~al.}}]{Daddi2007}
{Daddi}, E., {et~al.} 2007, \apj, 670, 156

\bibitem[{{Daddi} {et~al.}(2009){Daddi}, {Dannerbauer}, {Stern}, {Dickinson},
  {Morrison}, {Elbaz}, {Giavalisco}, {Mancini}, {Pope}, \&
  {Spinrad}}]{Daddi2009}
---. 2009, \apj, 694, 1517

\bibitem[{{Donnarumma} {et~al.}(2011){Donnarumma}, {Ettori}, {Meneghetti},
  {Gavazzi}, {Fort}, {Moscardini}, {Romano}, {Fu}, {Giordano}, {Radovich},
  {et~al.}}]{Donnarumma2011}
{Donnarumma}, A., {et~al.} 2011, \aap, 528, A73+

\bibitem[{{Duffy} {et~al.}(2008){Duffy}, {Schaye}, {Kay}, \& {Dalla
  Vecchia}}]{Duffy2008}
{Duffy}, A.~R., {Schaye}, J., {Kay}, S.~T., \& {Dalla Vecchia}, C. 2008,
  \mnras, 390, L64

\bibitem[{{Duffy} {et~al.}(2010){Duffy}, {Schaye}, {Kay}, {Dalla Vecchia},
  {Battye}, \& {Booth}}]{Duffy2010}
{Duffy}, A.~R., {Schaye}, J., {Kay}, S.~T., {Dalla Vecchia}, C., {Battye},
  R.~A., \& {Booth}, C.~M. 2010, \mnras, 405, 2161

\bibitem[{{Evrard} {et~al.}(2002){Evrard}, {MacFarland}, {Couchman}, {Colberg},
  {Yoshida}, {White}, {Jenkins}, {Frenk}, {Pearce}, {Peacock}, \&
  {Thomas}}]{Evrard2002}
{Evrard}, A.~E., {et~al.} 2002, \apj, 573, 7

\bibitem[{{Fedeli} {et~al.}(2010){Fedeli}, {Meneghetti}, {Gottl{\"o}ber}, \&
  {Yepes}}]{Fedeli2010}
{Fedeli}, C., {Meneghetti}, M., {Gottl{\"o}ber}, S., \& {Yepes}, G. 2010, \aap,
  519, A91+

\bibitem[{{Fioc} \& {Rocca-Volmerange}(1997)}]{Fioc1997PEGASE}
{Fioc}, M., \& {Rocca-Volmerange}, B. 1997, \aap, 326, 950

\bibitem[{{Gavazzi} {et~al.}(2003){Gavazzi}, {Fort}, {Mellier}, {Pell{\'o}}, \&
  {Dantel-Fort}}]{Gavazzi2003}
{Gavazzi}, R., {Fort}, B., {Mellier}, Y., {Pell{\'o}}, R., \& {Dantel-Fort}, M.
  2003, \aap, 403, 11

\bibitem[{{Gobat} {et~al.}(2011){Gobat}, {Daddi}, {Onodera}, {Finoguenov},
  {Renzini}, {Arimoto}, {Bouwens}, {Brusa}, {Chary}, {Cimatti},
  {et~al.}}]{Gobat2011}
{Gobat}, R., {et~al.} 2011, \aap, 526, A133+

\bibitem[{{Gottl{\"o}ber} \& {Yepes}(2007)}]{Gottlober2007}
{Gottl{\"o}ber}, S., \& {Yepes}, G. 2007, \apj, 664, 117

\bibitem[{{Grogin} {et~al.}(2010){Grogin}, {Lucas}, {Golimowski}, \&
  {Biretta}}]{Grogin2010}
{Grogin}, N.~A., {Lucas}, R., {Golimowski}, D., \& {Biretta}, J. 2010, {WFPC2
  CTE for Extended Sources: I. Photometric Correction}, Tech. rep.

\bibitem[{{Halkola} {et~al.}(2006){Halkola}, {Seitz}, \&
  {Pannella}}]{Halkola2006}
{Halkola}, A., {Seitz}, S., \& {Pannella}, M. 2006, \mnras, 372, 1425

\bibitem[{{Hennawi} {et~al.}(2007){Hennawi}, {Dalal}, {Bode}, \&
  {Ostriker}}]{Hennawi2007}
{Hennawi}, J.~F., {Dalal}, N., {Bode}, P., \& {Ostriker}, J.~P. 2007, \apj,
  654, 714

\bibitem[{{Hildebrandt} {et~al.}(2010){Hildebrandt}, {Arnouts}, {Capak},
  {Moustakas}, {Wolf}, {Abdalla}, {Assef}, {Banerji}, {Ben{\'{\i}}tez},
  {Brammer}, {et~al.}}]{Hildebrandt2010PHOTZ}
{Hildebrandt}, H., {et~al.} 2010, \aap, 523, A31+

\bibitem[{{Huang} {et~al.}(2011){Huang}, {Radovich}, {Grado}, {Puddu},
  {Romano}, {Limatola}, \& {Fu}}]{Huang2011}
{Huang}, Z., {Radovich}, M., {Grado}, A., {Puddu}, E., {Romano}, A.,
  {Limatola}, L., \& {Fu}, L. 2011, arXiv, 1102.1837

\bibitem[{{Ilbert} {et~al.}(2006){Ilbert}, {Arnouts}, {McCracken},
  {Bolzonella}, {Bertin}, {Le F{\`e}vre}, {Mellier}, {Zamorani}, {Pell{\`o}},
  {Iovino}, {et~al.}}]{Ilbert2006BPZ}
{Ilbert}, O., {et~al.} 2006, \aap, 457, 841

\bibitem[{{Ilbert} {et~al.}(2009){Ilbert}, {Capak}, {Salvato}, {Aussel},
  {McCracken}, {Sanders}, {Scoville}, {Kartaltepe}, {Arnouts}, {Le Floc'h},
  {et~al.}}]{Ilbert2009}
---. 2009, \apj, 690, 1236

\bibitem[{{Jee} {et~al.}(2009){Jee}, {Rosati}, {Ford}, {Dawson}, {Lidman},
  {Perlmutter}, {Demarco}, {Strazzullo}, {Mullis}, {B{\"o}hringer},
  {et~al.}}]{Jee2009}
{Jee}, M.~J., {et~al.} 2009, \apj, 704, 672

\bibitem[{{Jim{\'e}nez-Teja} \& {Ben{\'{\i}}tez}(2011)}]{Jimenez&Benitez2011}
{Jim{\'e}nez-Teja}, Y., \& {Ben{\'{\i}}tez}, N. 2011, arXiv, 1104.0683

\bibitem[{{Jing} \& {Suto}(2000)}]{JingSuto2000}
{Jing}, Y.~P., \& {Suto}, Y. 2000, \apjl, 529, L69

\bibitem[{{Jullo} {et~al.}(2010){Jullo}, {Natarajan}, {Kneib}, {D'Aloisio},
  {Limousin}, {Richard}, \& {Schimd}}]{Jullo2010}
{Jullo}, E., {Natarajan}, P., {Kneib}, J., {D'Aloisio}, A., {Limousin}, M.,
  {Richard}, J., \& {Schimd}, C. 2010, Science, 329, 924

\bibitem[{{Kaiser} \& {Squires}(1993)}]{KaiserSquires1993}
{Kaiser}, N., \& {Squires}, G. 1993, \apj, 404, 441

\bibitem[{{Klypin} {et~al.}(2010){Klypin}, {Trujillo-Gomez}, \&
  {Primack}}]{Klypin2010}
{Klypin}, A., {Trujillo-Gomez}, S., \& {Primack}, J. 2010, arXiv, 1002.3660

\bibitem[{{Koekemoer} {et~al.}(2007){Koekemoer}, {Aussel}, {Calzetti}, {Capak},
  {Giavalisco}, {Kneib}, {Leauthaud}, {Le F{\`e}vre}, {McCracken}, {Massey},
  {Mobasher}, {et~al.}}]{Koekemoer2007}
{Koekemoer}, A.~M., {et~al.} 2007, \apjs, 172, 196

\bibitem[{{Lahav} {et~al.}(1991){Lahav}, {Lilje}, {Primack}, \&
  {Rees}}]{Lahav1991}
{Lahav}, O., {Lilje}, P.~B., {Primack}, J.~R., \& {Rees}, M.~J. 1991, \mnras,
  251, 128

\bibitem[{{Lemze} {et~al.}(2009){Lemze}, {Sadeh}, \& {Rephaeli}}]{Lemze2009}
{Lemze}, D., {Sadeh}, S., \& {Rephaeli}, Y. 2009, \mnras, 397, 1876

\bibitem[{{Liesenborgs} {et~al.}(2006){Liesenborgs}, {De Rijcke}, \&
  {Dejonghe}}]{Liesenborgs2006}
{Liesenborgs}, J., {De Rijcke}, S., \& {Dejonghe}, H. 2006, \mnras, 367, 1209

\bibitem[{{Liesenborgs} {et~al.}(2007){Liesenborgs}, {de Rijcke}, {Dejonghe},
  \& {Bekaert}}]{Liesenborgs2007}
{Liesenborgs}, J., {de Rijcke}, S., {Dejonghe}, H., \& {Bekaert}, P. 2007,
  \mnras, 380, 1729

\bibitem[{{Liesenborgs} {et~al.}(2009){Liesenborgs}, {de Rijcke}, {Dejonghe},
  \& {Bekaert}}]{Liesenborgs2009}
---. 2009, \mnras, 397, 341

\bibitem[{{Limousin} {et~al.}(2008){Limousin}, {Richard}, {Kneib}, {Brink},
  {Pell{\'o}}, {Jullo}, {Tu}, {Sommer-Larsen}, {Egami}, {Micha{\l}owski},
  {et~al.}}]{Limousin2008}
{Limousin}, M., {et~al.} 2008, \aap, 489, 23

\bibitem[{{Medezinski} {et~al.}(2011){Medezinski}, {Broadhurst}, {Umetsu},
  {Benitez}, \& {Taylor}}]{Medezinski2011}
{Medezinski}, E., {Broadhurst}, T., {Umetsu}, K., {Benitez}, N., \& {Taylor},
  A. 2011, arXiv, 1101.1955

\bibitem[{{Medezinski} {et~al.}(2010){Medezinski}, {Broadhurst}, {Umetsu},
  {Oguri}, {Rephaeli}, \& {Ben{\'{\i}}tez}}]{Medezinski2010}
{Medezinski}, E., {Broadhurst}, T., {Umetsu}, K., {Oguri}, M., {Rephaeli}, Y.,
  \& {Ben{\'{\i}}tez}, N. 2010, \mnras, 405, 257

\bibitem[{{Meneghetti} {et~al.}(2003){Meneghetti}, {Bartelmann}, \&
  {Moscardini}}]{Meneghetti2003}
{Meneghetti}, M., {Bartelmann}, M., \& {Moscardini}, L. 2003, \mnras, 346, 67

\bibitem[{{Meneghetti} {et~al.}(2010{\natexlab{a}}){Meneghetti}, {Fedeli},
  {Pace}, {Gottl{\"o}ber}, \& {Yepes}}]{Meneghetti2010a}
{Meneghetti}, M., {Fedeli}, C., {Pace}, F., {Gottl{\"o}ber}, S., \& {Yepes}, G.
  2010{\natexlab{a}}, \aap, 519, A90+

\bibitem[{{Meneghetti} {et~al.}(2011){Meneghetti}, {Fedeli}, {Zitrin},
  {Bartelmann}, {Broadhurst}, {Gottloeber}, {Moscardini}, \&
  {Yepes}}]{Meneghetti2011}
{Meneghetti}, M., {Fedeli}, C., {Zitrin}, A., {Bartelmann}, M., {Broadhurst},
  T., {Gottloeber}, S., {Moscardini}, L., \& {Yepes}, G. 2011, arXiv, 1103.0044

\bibitem[{{Meneghetti} {et~al.}(2010{\natexlab{b}}){Meneghetti}, {Rasia},
  {Merten}, {Bellagamba}, {Ettori}, {Mazzotta}, {Dolag}, \&
  {Marri}}]{Meneghetti2010b}
{Meneghetti}, M., {Rasia}, E., {Merten}, J., {Bellagamba}, F., {Ettori}, S.,
  {Mazzotta}, P., {Dolag}, K., \& {Marri}, S. 2010{\natexlab{b}}, \aap, 514,
  A93+

\bibitem[{{Merten} {et~al.}(2009){Merten}, {Cacciato}, {Meneghetti}, {Mignone},
  \& {Bartelmann}}]{Merten2009}
{Merten}, J., {Cacciato}, M., {Meneghetti}, M., {Mignone}, C., \& {Bartelmann},
  M. 2009, \aap, 500, 681

\bibitem[{{Merten} {et~al.}(2011){Merten}, {Coe}, {Dupke}, {Massey}, {Zitrin},
  {Cypriano}, {Okabe}, {Frye}, {Braglia}, {Jimenez-Teja},
  {et~al.}}]{Merten2011}
{Merten}, J., {et~al.} 2011, arXiv, 1103.2772

\bibitem[{{Morandi} \& {Limousin}(2011)}]{Morandi2011On383}
{Morandi}, A., \& {Limousin}, M. 2011, arXiv, 1108.0769

\bibitem[{{Morandi} {et~al.}(2011){Morandi}, {Limousin}, {Rephaeli}, {Umetsu},
  {Barkana}, {Broadhurst}, \& {Dahle}}]{Morandi2011}
{Morandi}, A., {Limousin}, M., {Rephaeli}, Y., {Umetsu}, K., {Barkana}, R.,
  {Broadhurst}, T., \& {Dahle}, H. 2011, arXiv, 1103.0202

\bibitem[{{Navarro} {et~al.}(1996){Navarro}, {Frenk}, \& {White}}]{Navarro1996}
{Navarro}, J.~F., {Frenk}, C.~S., \& {White}, S.~D.~M. 1996, \apj, 462, 563

\bibitem[{{Neto} {et~al.}(2007){Neto}, {Gao}, {Bett}, {Cole}, {Navarro},
  {Frenk}, {White}, {Springel}, \& {Jenkins}}]{Neto2007}
{Neto}, A.~F., {et~al.} 2007, \mnras, 381, 1450

\bibitem[{{Newman} {et~al.}(2011){Newman}, {Treu}, {Ellis}, \&
  {Sand}}]{Newman2011}
{Newman}, A.~B., {Treu}, T., {Ellis}, R.~S., \& {Sand}, D.~J. 2011, \apjl, 728,
  L39+

\bibitem[{{Newman} {et~al.}(2009){Newman}, {Treu}, {Ellis}, {Sand}, {Richard},
  {Marshall}, {Capak}, \& {Miyazaki}}]{Newman2009}
{Newman}, A.~B., {Treu}, T., {Ellis}, R.~S., {Sand}, D.~J., {Richard}, J.,
  {Marshall}, P.~J., {Capak}, P., \& {Miyazaki}, S. 2009, \apj, 706, 1078

\bibitem[{{Oguri} \& {Blandford}(2009)}]{OguriBlandford2009}
{Oguri}, M., \& {Blandford}, R.~D. 2009, \mnras, 392, 930

\bibitem[{{Oguri} {et~al.}(2009){Oguri}, {Hennawi}, {Gladders}, {Dahle},
  {Natarajan}, {Dalal}, {Koester}, {Sharon}, \& {Bayliss}}]{Oguri2009}
{Oguri}, M., {et~al.} 2009, \apj, 699, 1038

\bibitem[{{Okabe} {et~al.}(2010){Okabe}, {Takada}, {Umetsu}, {Futamase}, \&
  {Smith}}]{Okabe2010}
{Okabe}, N., {Takada}, M., {Umetsu}, K., {Futamase}, T., \& {Smith}, G.~P.
  2010, \pasj, 62, 811

\bibitem[{{Peebles}(1985)}]{Peebles1985}
{Peebles}, P.~J.~E. 1985, \apj, 297, 350

\bibitem[{{Postman} {et~al.}(2011){Postman}, {Coe}, {Benitez}, {Bradley},
  {Broadhurst}, {Donahue}, {Ford}, {Graur}, {Graves}, {Jouvel},
  {et~al.}}]{Postman2011CLASHoverview}
{Postman}, M., {et~al.} 2011, ApJS submitted, arXiv, 1106.3328

\bibitem[{{Prada} {et~al.}(2011){Prada}, {Klypin}, {Cuesta}, {Betancort-Rijo},
  \& {Primack}}]{Prada2011SimCDM}
{Prada}, F., {Klypin}, A.~A., {Cuesta}, A.~J., {Betancort-Rijo}, J.~E., \&
  {Primack}, J. 2011, arXiv, 1104.5130

\bibitem[{{Richard} {et~al.}(2011){Richard}, {Kneib}, {Ebeling}, {Stark},
  {Egami}, \& {Fiedler}}]{Richard2011}
{Richard}, J., {Kneib}, J., {Ebeling}, H., {Stark}, D., {Egami}, E., \&
  {Fiedler}, A.~K. 2011, arXiv, 1102.5092

\bibitem[{{Rosati} {et~al.}(2009){Rosati}, {Tozzi}, {Gobat}, {Santos},
  {Nonino}, {Demarco}, {Lidman}, {Mullis}, {Strazzullo}, {B{\"o}hringer},
  {et~al.}}]{Rosati2009}
{Rosati}, P., {et~al.} 2009, \aap, 508, 583

\bibitem[{{Sadeh} \& {Rephaeli}(2008)}]{SadehRephaeli2008}
{Sadeh}, S., \& {Rephaeli}, Y. 2008, \mnras, 388, 1759

\bibitem[{{Sand} {et~al.}(2008){Sand}, {Treu}, {Ellis}, {Smith}, \&
  {Kneib}}]{Sand2008}
{Sand}, D.~J., {Treu}, T., {Ellis}, R.~S., {Smith}, G.~P., \& {Kneib}, J. 2008,
  \apj, 674, 711

\bibitem[{{Sand} {et~al.}(2004){Sand}, {Treu}, {Smith}, \& {Ellis}}]{Sand2004}
{Sand}, D.~J., {Treu}, T., {Smith}, G.~P., \& {Ellis}, R.~S. 2004, \apj, 604,
  88

\bibitem[{{Schmidt} \& {Allen}(2007)}]{SchmidtAllen2007}
{Schmidt}, R.~W., \& {Allen}, S.~W. 2007, \mnras, 379, 209

\bibitem[{{Sereno} {et~al.}(2010){Sereno}, {Jetzer}, \& {Lubini}}]{Sereno2010}
{Sereno}, M., {Jetzer}, P., \& {Lubini}, M. 2010, \mnras, 403, 2077

\bibitem[{{Smith} {et~al.}(2001){Smith}, {Kneib}, {Ebeling}, {Czoske}, \&
  {Smail}}]{Smith2001}
{Smith}, G.~P., {Kneib}, J., {Ebeling}, H., {Czoske}, O., \& {Smail}, I. 2001,
  \apj, 552, 493

\bibitem[{{Smith} {et~al.}(2005){Smith}, {Kneib}, {Smail}, {Mazzotta},
  {Ebeling}, \& {Czoske}}]{Smith2005}
{Smith}, G.~P., {Kneib}, J., {Smail}, I., {Mazzotta}, P., {Ebeling}, H., \&
  {Czoske}, O. 2005, \mnras, 359, 417

\bibitem[{{Umetsu} \& {Broadhurst}(2008)}]{UmetsuBroadhurst2008}
{Umetsu}, K., \& {Broadhurst}, T. 2008, \apj, 684, 177

\bibitem[{{Umetsu} {et~al.}(2011{\natexlab{a}}){Umetsu}, {Broadhurst},
  {Zitrin}, {Medezinski}, {Coe}, \& {Postman}}]{Umetsu2011b}
{Umetsu}, K., {Broadhurst}, T., {Zitrin}, A., {Medezinski}, E., {Coe}, D., \&
  {Postman}, M. 2011{\natexlab{a}}, arXiv, 1105.0444

\bibitem[{{Umetsu} {et~al.}(2011{\natexlab{b}}){Umetsu}, {Broadhurst},
  {Zitrin}, {Medezinski}, \& {Hsu}}]{Umetsu2011a}
{Umetsu}, K., {Broadhurst}, T., {Zitrin}, A., {Medezinski}, E., \& {Hsu}, L.
  2011{\natexlab{b}}, \apj, 729, 127

\bibitem[{{Umetsu} {et~al.}(2010){Umetsu}, {Medezinski}, {Broadhurst},
  {Zitrin}, {Okabe}, {Hsieh}, \& {Molnar}}]{Umetsu2010}
{Umetsu}, K., {Medezinski}, E., {Broadhurst}, T., {Zitrin}, A., {Okabe}, N.,
  {Hsieh}, B., \& {Molnar}, S.~M. 2010, \apj, 714, 1470

\bibitem[{{Wuyts} {et~al.}(2008){Wuyts}, {Labb{\'e}}, {Schreiber}, {Franx},
  {Rudnick}, {Brammer}, \& {van Dokkum}}]{Wuyts2008PHOTZ}
{Wuyts}, S., {Labb{\'e}}, I., {Schreiber}, N.~M.~F., {Franx}, M., {Rudnick},
  G., {Brammer}, G.~B., \& {van Dokkum}, P.~G. 2008, \apj, 689, 653

\bibitem[{{Zhao} {et~al.}(2003){Zhao}, {Jing}, {Mo}, \&
  {B{\"o}rner}}]{Zhao2003}
{Zhao}, D.~H., {Jing}, Y.~P., {Mo}, H.~J., \& {B{\"o}rner}, G. 2003, \apjl,
  597, L9

\bibitem[{{Zhao} {et~al.}(2009){Zhao}, {Jing}, {Mo}, \&
  {B{\"o}rner}}]{Zhao2009}
---. 2009, \apj, 707, 354

\bibitem[{{Zhao}(1996)}]{Zhao1996}
{Zhao}, H. 1996, \mnras, 278, 488

\bibitem[{{Zitrin} \& {Broadhurst}(2009)}]{ZitrinBroadhurst2009}
{Zitrin}, A., \& {Broadhurst}, T. 2009, \apjl, 703, L132

\bibitem[{{Zitrin} {et~al.}(2011{\natexlab{a}}){Zitrin}, {Broadhurst},
  {Barkana}, {Rephaeli}, \& {Ben{\'{\i}}tez}}]{Zitrin2011a}
{Zitrin}, A., {Broadhurst}, T., {Barkana}, R., {Rephaeli}, Y., \&
  {Ben{\'{\i}}tez}, N. 2011{\natexlab{a}}, \mnras, 410, 1939

\bibitem[{{Zitrin} {et~al.}(2011{\natexlab{b}}){Zitrin}, {Broadhurst},
  {Bartelmann}, {Rephaeli}, {Oguri}, {Ben{\'{\i}}tez}, {Hao}, \&
  {Umetsu}}]{Zitrin2011c}
{Zitrin}, A., {Broadhurst}, T., {Bartelmann}, M., {Rephaeli}, Y., {Oguri}, M.,
  {Ben{\'{\i}}tez}, N., {Hao}, J., \& {Umetsu}, K. 2011{\natexlab{b}}, arXiv,
  1105.2295

\bibitem[{{Zitrin} {et~al.}(2011{\natexlab{c}}){Zitrin}, {Broadhurst}, {Coe},
  {Liesenborgs}, {Ben{\'{\i}}tez}, {Rephaeli}, {Ford}, \&
  {Umetsu}}]{Zitrin2011b}
{Zitrin}, A., {Broadhurst}, T., {Coe}, D., {Liesenborgs}, J., {Ben{\'{\i}}tez},
  N., {Rephaeli}, Y., {Ford}, H., \& {Umetsu}, K. 2011{\natexlab{c}}, \mnras,
  413, 1753

\bibitem[{{Zitrin} {et~al.}(2009{\natexlab{a}}){Zitrin}, {Broadhurst},
  {Rephaeli}, \& {Sadeh}}]{Zitrin2009a}
{Zitrin}, A., {Broadhurst}, T., {Rephaeli}, Y., \& {Sadeh}, S.
  2009{\natexlab{a}}, \apjl, 707, L102

\bibitem[{{Zitrin} {et~al.}(2009{\natexlab{b}}){Zitrin}, {Broadhurst},
  {Umetsu}, {Coe}, {Ben{\'{\i}}tez}, {Ascaso}, {Bradley}, {Ford}, {Jee},
  {Medezinski}, {et~al.}}]{Zitrin2009b}
{Zitrin}, A., {et~al.} 2009{\natexlab{b}}, \mnras, 396, 1985

\bibitem[{{Zitrin} {et~al.}(2010){Zitrin}, {Broadhurst}, {Umetsu}, {Rephaeli},
  {Medezinski}, {Bradley}, {Jim{\'e}nez-Teja}, {Ben{\'{\i}}tez}, {Ford},
  {Liesenborgs}, {de Rijcke}, {et~al.}}]{Zitrin2010}
---. 2010, \mnras, 408, 1916

\end{thebibliography}

\end{document}